\documentclass[twocolumn,prd,nofootinbib,aps,prl,floats,floatfix,amsmath,
amssymb,secnumarabic,aas_macros]{revtex4-1}

\usepackage[english]{babel}
\usepackage[utf8]{inputenc}
\usepackage[final]{graphicx}
\usepackage{hyperref}
\usepackage{amsmath}
\usepackage{bbm}
\usepackage{amsfonts}
\usepackage{amssymb}
\usepackage{latexsym}
\usepackage{graphicx}

\usepackage{multirow}
\usepackage{float}
\usepackage{url}
\usepackage{slashed}
\usepackage{xcolor}

\usepackage{cprotect}
\usepackage{verbatim}

\graphicspath{ {figures/} }

\newcommand{\be}{\begin{equation}}
\newcommand{\ee}{\end{equation}}
\newcommand{\ba}{\begin{array}}
\newcommand{\ea}{\end{array}}
\newcommand{\bea}{\begin{eqnarray}}
\newcommand{\eea}{\end{eqnarray}}
\newcommand{\sss}{\scriptscriptstyle}

\newcommand{\nn}{\nonumber}

\newcommand{\ob}{\Omega_b}
\newcommand{\obp}{\Omega_{b'}}
\newcommand{\he}{_{\mathrm{He}}}
\newcommand{\hei}{_{\mathrm{He^0}}}
\newcommand{\heii}{_{\mathrm{He^+}}}
\newcommand{\heiii}{_{\mathrm{He^{++}}}}
\newcommand{\h}{_{\mathrm{H}}}
\newcommand{\hh}{_{\mathrm{H_2}}}
\newcommand{\hi}{_{\mathrm{H^0}}}
\newcommand{\hii}{_{\mathrm{H^+}}}
\newcommand{\hm}{_{\mathrm{H}^-}}
\newcommand{\mn}{\overline{m_N}}
\newcommand{\cdm}{_{c}}
\newcommand{\vir}{_{\mathrm{vir}}}
\newcommand{\cool}{\mathcal{C}}
\DeclareRobustCommand{\MS}{\reflectbox{SM}}

\makeatletter
\def\apjs{\ref@jnl{ApJS}}               % Astrophysical Journal, Supplement
\makeatother 

\renewcommand{\arraystretch}{1.5}

\begin{document}

\title{Constraining galactic structures of mirror dark matter}

\author{Jean-Samuel Roux}
\author{James M.\ Cline}
\affiliation{McGill University, Department of Physics, 3600 University St.,
Montr\'eal, QC H3A2T8 Canada}
\begin{abstract}
The simplest model of mirror sector dark matter maintains exact mirror
symmetry, but has a baryon abundance $\Omega_{b'} = \beta \Omega_b$ and a 
suppressed temperature $T' = x T$ in
the mirror sector; hence it depends only on two parameters, $\beta,x$. 
For sufficiently small $x$, early cosmological observables may not
constrain mirror baryons from constituting all of the
dark matter despite their strong self-interactions, depending on the
unknown details of structure formation in the hidden sector.  
Here we close this loophole by simulating mirror structure formation,
mapping out the allowed regions of parameter space using cosmological and 
astronomical data. We find that the Milky Way disk surface density and bulge 
mass constrain $\Omega_{b'}\lesssim 0.3 \Omega_{b}$ at the highest 
$T'$ allowed by BBN and CMB ($T'=0.5 T$), or $\Omega_{b'}\lesssim 0.8 \Omega_{b}$ at 
lower values of $T'$. We also briefly discuss the realization of the necessary 
temperature asymmetry between the SM and the mirror sector in our model with 
unbroken mirror symmetry.

\end{abstract}
\maketitle

\section{Introduction}

The idea of a mirror copy of the standard model is the earliest
example of the now-popular paradigm of hidden dark sectors 
\cite{Kobzarev:1966qya,Okun:2006eb}.  The model has a number of
appealing features, the most obvious being that a CP-like transformation
is restored as a symmetry of 
nature, and dark matter candidates (mirror baryons) are provided 
\cite{Hodges:1993yb},
including a resolution of the cusp-core problem
\cite{Mohapatra:2000qx}. Twin Higgs models are one specific 
realization 
of mirror matter set to solve the little hierarchy problem 
\cite{Chacko:2005pe}.  More recently a subdominant mirror sector 
(\MS) has been suggested for stimulating the early growth of supermassive
black holes \cite{DAmico:2017lqj,Latif:2018kqv} and neutron -- mirror 
neutron oscillations have been proposed as a solution to the neutron lifetime 
puzzle \cite{Berezhiani:2018eds}. A detailed review of the general 
model is available in ref.\ \cite{Foot:2014mia}.

{\it A priori,} it seems possible that mirror baryons could constitute
all of the dark matter (DM), even though dark atoms have an interaction
cross section far exceeding the bounds from the Bullet Cluster
\cite{Markevitch:2003at,Randall:2007ph}.  If the mirror baryons are
present primarily in collapsed structures rather than gaseous atoms or
molecules, their self-interaction cross section would be sufficiently
small, just like ordinary stars are effectively collisionless. 
Moreover the strong constraints on atomic \cite{Cyr-Racine:2013fsa} or 
mirror \cite{Ciarcelluti:2010zz} dark matter from dark acoustic oscillations 
can be evaded if the
\MS\ temperature $T'$ is sufficiently low ($\lesssim 0.3\,T$)
compared to that of visible photons.  

A major goal of the present work is to determine what fraction of the
total DM density could be in mirror particles, by studying structure
formation in the \MS.  Assuming that mirror symmetry is
unbroken, we can do this exhaustively, since there are only two
parameters to vary: the relative abundance of mirror versus visible
baryons $\beta = \Omega_{b'}/\Omega_b$, and the temperature ratio
$x=T'/T$.   An additional particle that is noninteracting and
uncharged under the mirror symmetry is taken to comprise the remainder
of the DM, if necessary. Throughout this work, primes will distinguish elements of
the \MS\ analogous to the visible ones.  We assume that the
possible portal interactions between the two sectors (Higgs mixing
$h^2 h'^2$ and gauge kinetic mixing $F^{\mu\nu}F'_{\mu\nu}$) are
negligible, since these would cause $x\to 1$ if they were sufficiently
strong.

We adopt a methodology similar to ref.\ \cite{Ghalsasi:2017jna}, which studied
structure formation in a simplified atomic dark matter model.  Namely
we use the extended Press-Schechter formalism
\cite{Press:1973iz,Bond:1990iw,Lacey:1993iv} and the semi-analytical model 
GALFORM \cite{Cole:2000ex} to simulate the merger history of DM halos and study 
the formation of dark galactic structures. Unlike the dark atomic model, 
mirror DM contains nuclear reactions which allow mirror helium formation and 
stellar feedback to alter the evolution of DM structures. We consider the 
effects of these extra features in the hidden sector quantitatively.

In order to predict structure formation, one must first understand the
early-universe cosmology of the model, leading to the primordial
mirror abundances and ionization fractions.  As well, constraints
on additional radiation degrees of freedom are imposed by the cosmic
microwave background (CMB) and big-bang nucleosynthesis (BBN).
This is worked out in 
sect.\ \ref{sect:cosmo}.  These inputs allow us to simulate structure 
formation in the \MS\ using the semi-analytical galaxy formation model GALFORM, 
which we describe in sect.\ \ref{sec:sf}. In sect.\ \ref{sec:constraints}, 
we present our results and the constraints on the parameters $x,\beta$ 
coming from astronomical observations. In sect.\ \ref{sect:early} we 
discuss early cosmological scenarios that could produce values of $x,\beta$ 
consistent with our constraints without explicitly breaking the mirror 
symmetry.  Conclusions are given in sect.\ \ref{sect:conc}.

Throughout this paper we will use the following cosmological parameters 
\cite{Tanabashi:2018oca}: 
$h= 0.678$, $T_0=2.7255$~K, $\Omega_m = 0.308$, $\Omega_b = 0.0484$, 
$\Omega_\Lambda=0.692$, $n_s = 0.968$ and $\sigma_8 =0.815$. Although most of 
these values were derived assuming a $\Lambda$CDM cosmology, our conclusions 
would not change significantly if we used slightly different parameters.

\section{Cosmology of the mirror sector}
\label{sect:cosmo}

We assume that a fraction of dark matter resides in a hidden sector whose
gauge group $G'$
is a copy of the Standard Model (SM) gauge group $G = SU(3) \times SU(2)\times 
U(1)$. This model possesses a discrete mirror symmetry $P_{G\leftrightarrow 
G'}$ that interchanges the 
fields of the ordinary, observable sector with their mirror counterparts. If 
$P_{G\leftrightarrow 
G'}$ is unbroken as we assume, then the microphysics of each sector is 
identical. In particular, mirror matter comprises the same chemical and nuclear 
species as ordinary matter and all their processes have the same rates. 
Although $P_{G\leftrightarrow G'}$ does not forbid the 
two renomalizable gauge kinetic mixing and Higgs portal interactions between the 
two sectors, we assume that the portal
couplings are sufficiently small as to have negligible impact on early cosmology
and structure formation. Hence SM particles interact with their mirror 
counterparts only gravitationally.

Eventually we will confirm that mirror baryons cannot comprise all of the DM,
necessitating an additional component in the form of standard 
cold, collisionless dark matter (CDM) that is assumed to
interact with the 
other sectors only gravitationally. The total matter density in the universe 
is then $\Omega_m = \Omega_{c} + \ob + \obp$,
where $\Omega_{c}$ is the CDM fraction and $\ob$ ($\obp$) is the ordinary 
(mirror) baryon relic density, in units of the critical density. 

Because of the mirror symmetry,  $\ob$ and $\obp$ would likely
originate from the same mechanism; nevertheless  
$\obp$ can be different from $\ob$ if the two sectors have 
different initial temperatures 
\cite{1985Natur.314..415K,Berezhiani:2000gw}.  We accordingly take $\beta 
\equiv \obp / \ob$ as 
a second free parameter, in addition to the temperature ratio  $x \equiv T'/T$.
In fact $T'/T$ is time-dependent during early cosmology, for example through $e 
\bar{e}$ and $e' \bar{e}'$ annihilations occurring at different redshifts, 
which produces a relative difference of entropy in the photon backgrounds 
during some period.  But it becomes constant at the late times relevant for 
structure formation, hence we define $x$ to be the asymptotic value.  This 
remains true even in the presence of portal interactions between the two 
sectors, as long as they are weak enough to freeze out before the onset of 
structure formation.

\subsection{Effective radiation species}

BBN and the CMB constrain the expansion rate of the universe and thereby the total
radiative energy density. This is conventionally expressed as a limit on the number of 
additional effective neutrino species $\Delta N_{\textrm{eff}}= 
N_{\textrm{eff}}-3.046$. The contribution to $\Delta N_{\textrm{eff}}$ 
from the mirror photons and neutrinos follows from \cite{Berezhiani:2000gw}
\be  \label{eq:rhoRad}
\Delta \rho_{\textrm{rad}} = \frac{7}{8} \left( \frac{T_\nu}{T}\right)^{4} 
\Delta N_{\textrm{eff}} \  \rho_\gamma = \frac{\pi^2}{30} g_*' (T')\ T'^4, 
\ee 
where $\rho_\gamma$ is the energy density of ordinary photons and $g_*'(T')$ is 
the effective number of relativistic degrees of freedom in the \MS. 
Using $\rho_\gamma = (2\pi^2/30) T^4$ and $T_\nu/T = (4/11)^{1/3}$ today we find
that
\be 
\Delta N_{\textrm{eff}} = \frac{4}{7} \left( \frac{11}{4}\right)^{4/3} g_*' 
(T')\ x^4 .
\label{eq:DNeff}
\ee

The most recent data from the Planck Collaboration indicates that at the epoch 
of recombination $N_{\textrm{eff}} = 2.99^{+0.34}_{-0.33}$ with 95~\% 
confidence
\cite{Aghanim:2018eyx}, which gives the 3$\sigma$ limit $\Delta 
N_{\textrm{eff}}[\textrm{CMB}]<0.45$. At this temperature only photons 
and neutrinos are relativistic so $g_*'(T')= 3.38$, leading to the bound
\be 
x \lesssim 0.5. \text{~~~ (CMB)}\label{eq:xbound}
\ee 

BBN sets a similar  limit on $x$, even using the more stringent 
bound on the effective neutrino species $\Delta 
N_{\textrm{eff}} [\textrm{BBN}] \lesssim 0.3$ 
\cite{Cyburt:2015mya,Hufnagel:2017dgo}. 
This is because, although neutrinos are decoupled from photons at the BBN
temperature, nominally $T_{\sss BBN}\sim 1\,$MeV,  $e\bar{e}$ pairs 
haven't annihilated yet, such that $T_\nu \simeq T_\gamma$ and the factor of 
$(4/11)^{4/3}$ is removed from from eq.\ (\ref{eq:DNeff}).  Moreover for 
$x=0.5$, $g_*'(T') = g_*(0.5\,T_{\sss BBN})
\simeq 10$, leading to the bound $x < 0.48$, which is essentially the same as 
the CMB constraint (\ref{eq:xbound}). Using lower values of $g_*'(T')$ 
would make this limit less stringent.

\subsection{Mirror nucleosynthesis}

\begin{figure}[t]
\begin{center}
\includegraphics[width=\linewidth]{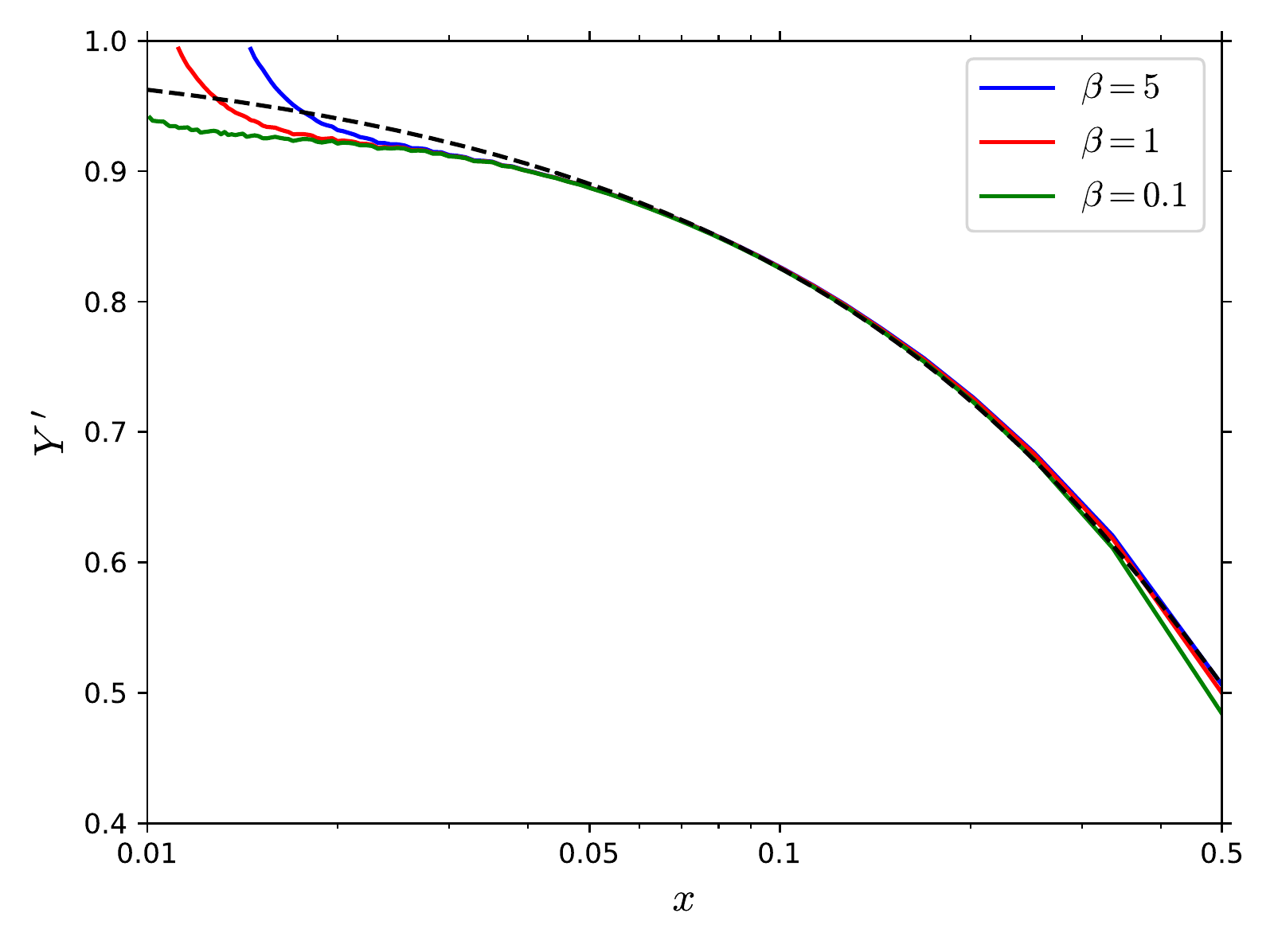}
\cprotect\caption{Mirror $^4$He relic abundances. The solid curves were 
computed numerically using \verb|AlterBBN| and the dashed line shows the
 approximate formula of eq.\ (\ref{eq:Hefrac}).}
\label{fig:helium}
\end{center} 
\end{figure}

The upper limit on $x$ leads to an important feature of mirror matter: it has a 
large helium abundance, stemming from the early freeze-out of  $n'\leftrightarrow p'$
equilibrium.  This implies a higher relic neutron abundance and consequently  more
efficient deuterium and helium production in the \MS.
Early freeze-out of mirror interactions is a general consequence of the temperature
hierarchy $T'<T$, which causes a cosmological event (tied to some temperature scale),
occurring at redshift  $z$ in the visible sector, to transpire at
higher redshift $z' \simeq -1 + (1+z)/x$ in the \MS.
Since the universe was expanding more rapidly at redshift $z'$ than at $z$,
the freeze-out of mirror processes will generally occur even earlier than
this estimated $z'$.

 An approximate formula for the relic $^4$He$'$ 
mass fraction was derived in ref.\ \cite{Berezhiani:2000gw},
\be 
Y'  \simeq \frac{2 \exp{\left[ -t_N/\tau(1+x^{-4})^{1/2}  \right]}}{1 + 
\exp{\left[\Delta m/T_W (1+x^{-4})^{1/6 }\right]}}, \label{eq:Hefrac}
\ee 
where $t_N\sim 200\,$s is the age of the universe at the ``deuterium 
bottleneck'', $\tau= 886.7\,$ s is the neutron lifetime, $T_W\simeq0.8\,$ MeV 
is the $n'\leftrightarrow p'$ freeze-out temperature, and $\Delta m = 1.29\,$ MeV 
is the neutron-proton mass difference. This relation is plotted in fig.\  
\ref{fig:helium}. 

The approximation (\ref{eq:Hefrac}) neglects possible dependence on $\beta$. 
A more accurate 
treatment of BBN is required to determine how the density of mirror baryons affects 
the freeze-out temperature of light nuclei and their relic abundances. We used 
the code \verb|AlterBBN| \cite{Arbey:2011nf,Arbey:2018zfh} to numerically 
compute the residual (mirror) $^4$He$'$ mass fraction for different values of $\beta$, 
modifying parameters of 
the code to match the conditions of the \MS. 
Namely the current CMB temperature, the baryon 
density and the baryon-to-photon ratio were replaced by
$T_0\rightarrow T_0/x$, $\ob\rightarrow \beta\ob$ and  $\eta_b 
\rightarrow 
\eta_b \beta/x^3$. Visible sector photons and neutrinos were incorporated
as additional effective 
neutrino species. Rewriting eqs.\ (\ref{eq:rhoRad},\ref{eq:DNeff}) from the perspective of the mirror world yields $\Delta 
N_{\textrm{eff}}' \simeq 7.44/x^4$ today. 

The results of our numerical calculations are also plotted in fig.\ 
\ref{fig:helium} for three benchmark values of $\beta$. Eq.\ (\ref{eq:Hefrac}) 
agrees with the numerical calculations within a few percent, which is sufficient
for our purposes in the following analysis. The striking domination of 
$^4$He$'$ in the \MS, in the limit
when $x\rightarrow0$ for fixed $\beta$, is contrary to statements made in 
refs.\ \cite{DAmico:2017lqj,Latif:2018kqv}.\footnote{These stem from
	a misinterpretation of ref.\ \cite{Berezhiani:2000gw}, which states that
	$Y'\to 0$ as $x\to 0$ with $\eta_{b'}$ fixed.  But this implies that $\beta$
	is varying with $x$, rather than being held fixed.}  One sees that for any 
	value of
$\beta$, the $^4$He$'$ fraction reaches $1$ as $x$ decreases to some critical 
value.  The \verb|AlterBBN| code is not suited to handle the situation where the
mirror hydrogen (H$'$) abundance vanishes since H$'$ is used as a reference to 
normalize other abundances.  Thus we cannot keep track of very small H$'$ 
densities. Moreover the age of the universe at the formation of mirror 
deuterium scales roughly as $t_N\sim x^2$; hence for small values of $x$ 
nucleosynthesis occurs in a fraction of a second and the Boltzmann equations 
for each species become too stiff for \verb|AlterBBN| to maintain a high 
accuracy. But this has no impact on our main results since both eq.\ 
(\ref{eq:Hefrac}) and our numerical analysis agree that the $^4$He$'$ 
abundance is limited to $0.9< Y' < 1$ for small values of $x$, with little 
phenomenological variation within this range.

The $^4$He$'$ abundance determines a number of other quantities that will 
be useful in the subsequent analysis.
Let $X_i\equiv n_i/n_{\mathrm{H}}$ be the 
relative abundance of a given chemical species, conventionally normalized to 
the H$'$ density\footnote{In what follows, we will drop the prime from H$'$ and 
H will refer to mirror hydrogen 
in all its chemical forms whereas H$^0$, H$^+$ and H$_2$ designate its neutral, 
ionized and molecular states respectively. Thus for a gas of pure H$_2$, $n\h= 
2 n\hh$. Similarly, He refers to all forms of mirror helium.}. The helium-hydrogen 
number ratio is 
\be 
X\he \equiv \frac{n\he}{n\h} = \frac{m_p}{m\he} \frac{Y'}{1-Y'} \simeq 
\frac{1}{4} \frac{Y'}{1-Y'}, \label{eq:xhe}
\ee
where $m_p$ is the proton mass. Furthermore the helium number fraction 
(distinct from the mass fraction $Y'= m\he n\he/(m\he n\he + m\h n\h) $)
is 
\be
\qquad\qquad f\he \equiv \frac{n\he}{n_N} = \frac{1}{1+ 1/X\he}\simeq 
\frac{Y'}{4-3Y'}, \qquad\ 
\label{eq:fhe}
\ee
\noindent with $n_N = n\h + n\he$ denoting the number density of nuclei. The 
mean mass per nucleus is
\begin{align} %align prevents overlap of the equation with the text
\mn = \left(\frac{1-Y'}{m_p} + \frac{Y'}{m\he} \right)^{-1} \simeq 
\frac{m_p}{1-\frac{3}{4}Y'}.  \label{eq:mn}
\end{align}

By virtue of the approximation made in eq.\ (\ref{eq:Hefrac}),
the expressions (\ref{eq:xhe}-\ref{eq:mn}) are independent of $\beta$. 
Lastly, the background number density of nuclei at any redshift follows 
from
\be \label{eq:nb}
n_N = \frac{3 H_0^2 \ob}{8 \pi G} \frac{\beta}{\mn}(1+z)^3 \ , 
\ee
implying that $n\h = (1-f\he)n_N$ and $n\he = f\he n_N$.

\subsection{Recombination}
\label{sect:recomb}
Due to its lower temperature, recombination in the \MS\ occurs at the higher
redshift $z_{\mathrm{rec}}'\simeq 
1100/x$. With its large fraction of mirror He, that has  a higher 
binding energy than H, this leads to more efficient recombination and a
lower residual free electron density. 
Primordial gas clouds require free electrons to cool and collapse into 
compact structures. Therefore the small relic ionization fraction has a 
direct impact on the formation of the first mirror stars.

Recombination proceeds through three major steps, which are the respective 
formation of 
He$^{+}$, He$^0$ and H$^0$. The latter is prevalent in the SM, but  
recombination of He is  more important in the \MS\ because of its  
high abundance. We adopt the effective three-level calculation presented 
in ref.\ \cite{Seager:1999bc} which was also used in ref.\ \cite{DAmico:2017lqj}. 

Recombination of He$^+$ in the SM occurred around $z\simeq6000$ 
($kT\sim1.4\,\mathrm{eV}$), at a sufficiently high density  for the 
ionized 
species to closely track their thermodynamic equilibrium abundances, in 
accordance with the Saha equation.  Since recombination occurs even 
earlier in the \MS, this is also true for mirror He$^+$. Its
Saha equation is
\be
\frac{\left(X_e-1-X\he\right) X_e}{1+2 X\he-X_e}=\frac{\left(2 \pi m_e k 
T'\right)^{3 / 2}}{h^{3} n\h} e^{-\chi\heii / k T'},
\label{eq:saha}
\ee
where $\chi\heii=54.4\ $eV is the He$^+$ ionization energy, $T' = 
x T_0 (1+z)$ is the mirror photon temperature and the free electron
ratio is 
$X_e = X\hii + X\heii + 2X\heiii$  from matter neutrality. For $kT'\sim 
1.4\,\mathrm{eV}$, the exponential in eq.\ (\ref{eq:saha})
is negligible, giving $X_e = 1+X\he$.  Eliminating $X_e$ and using the
fact that $X\hii\simeq 1$ and $X\hei \simeq 0$ (there is essentially no neutral
H or He until $T'$ falls below $\sim 10~\%$ of the $n=2$ ionization energies of
H or He, i.e. until $kT'\lesssim 0.4\,$eV), this implies $X\heiii\simeq 0$:
we can neglect any residual He$^{++}$ fraction, and both H and He are singly
ionized.

At later times, the evolution of the ionized states follows the network
\cite{Seager:1999bc}
\begin{widetext}
\bea
\label{eq:recHii}
\frac{dX\hii}{dz} &=&\frac{ \left(X_e X\hii n\h \alpha\h - \beta\h (1-X\hii)
e^{-h\nu\h/kT_M'}\right)%\\
\left(1+K\h \Lambda\h n\h (1-X\hii)\right)}{H(z)(1+z)(1+K\h(\Lambda\h+\beta\h)n\h(1-X\hii))},\\
\label{eq:recHeii}
\frac{dX\heii}{dz} &=& \frac{\left(X_e X\heii n\h \alpha\hei - \beta\hei (X\he-X\heii)
e^{-h\nu\he/kT_M'}\right) %\\
\left(  1+K\hei \Lambda\he n\h 
(X\he-X\heii)e^{-h\tilde{\nu}\he/kT_M'}\right)}{H(z)(1+z)(1+K\hei(\Lambda\he+\beta\hei)n\h(X\he-X\heii)e^{-h\tilde{\nu}\he/kT_M'})},
\\
\label{eq:recTm}
\frac{dT_M'}{dz} &=& \frac{8 \sigma_T a_R T'^4}{3 H(z)(1+z)m_e c} \left(\frac{X_e}{1+X\he+X_e}\right) (T_M'-T') + \frac{2T_M'}{(1+z)}.
\eea
\end{widetext}
that describes the  the evolution of the He$^+$ and H$^+$
fractions, and the matter temperature $T_M'$, which at low redshifts is below 
the radiation temperature $T'$. The various parameters are specified in 
appendix \ref{appA}.
 
We used \verb|Recfast++| \cite{Seager:1999bc,Chluba:2010ca,RubinoMartin:2009ry,
Chluba:2010fy,Switzer:2007sn,Grin:2009ik,2010PhRvD..82f3521A} to solve the
system (\ref{eq:recHii}-\ref{eq:recTm}), making the same modifications as for
 \verb|AlterBBN|, with the He mass fraction given by 
eq.\ (\ref{eq:Hefrac}). It was assumed that the species were initially singly ionized 
($X\hii=1$, $X\heii=X\he$) and that matter was strongly coupled to radiation 
($T_M'=T'$). The initial redshift was taken to be 
sufficiently high  to encompass the beginning of H$^0$ and He$^0$ recombination,
and the system was evolved until $z=10$, the initial redshift of 
the subsequent structure formation analysis (see below).
Fig.\ \ref{fig:recEvol} shows the resulting evolution of the free electron fraction 
$f_e' = n_e/n_{e,\mathrm{tot}}$ (where $n_{e,\mathrm{tot}}$ includes the electrons in 
the ground state of He$^+$) during recombination for several values of 
$x$ (differentiated by color) and $\beta$ (differentiated by linestyle). 
The expected $x$-dependence of the redshift of recombination $z_{\rm rec}'\sim1100/x$
is evident, scaling inversely to the \MS\ temperature.

\begin{figure}[b]
		\includegraphics[width=\linewidth]{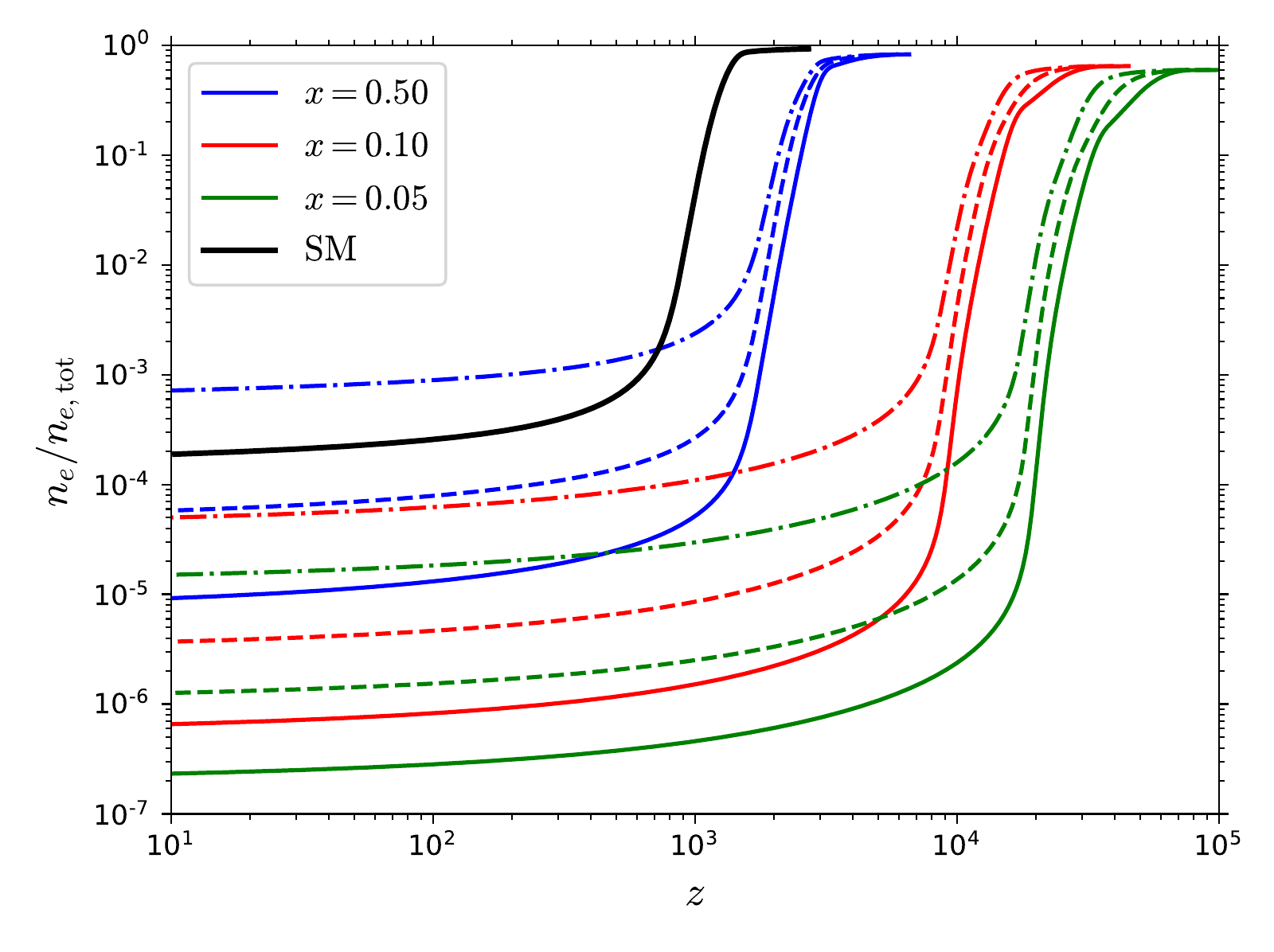}
		\cprotect\caption{ Evolution of the total ionization fraction during 
		mirror recombination. The solid, dashed and dash-dotted curves represent
		$\beta=$ 5, 1 and 0.1, respectively. Also	shown for comparison is 
		recombination in the SM.}
		\label{fig:recEvol}
\end{figure}

The most important feature for structure formation is the residual ionization 
fraction $f_e'$ at low redshifts. As fig.\ \ref{fig:recEvol} demonstrates, 
$f_e'$ is typically much smaller in the \MS\ than in the SM ($f_e\sim 2\times 
10^{-4}$). 
Only when $\beta\ll 1$  can $f_e'$ reach higher values, 
because the low density reduces the number of ion-electron 
collisions and the overall efficiency of recombination. But in this case the 
total electron density is also suppressed by a factor of $\beta$, so
the free electron density after recombination is always smaller in the 
\MS. This can have important consequences for early structure 
formation, since without a significant ionization fraction mirror matter clouds 
may not cool and collapse to form structures like ordinary matter does.

\begin{figure*}[t]
	\begin{center}
		\includegraphics[width=0.5\linewidth]{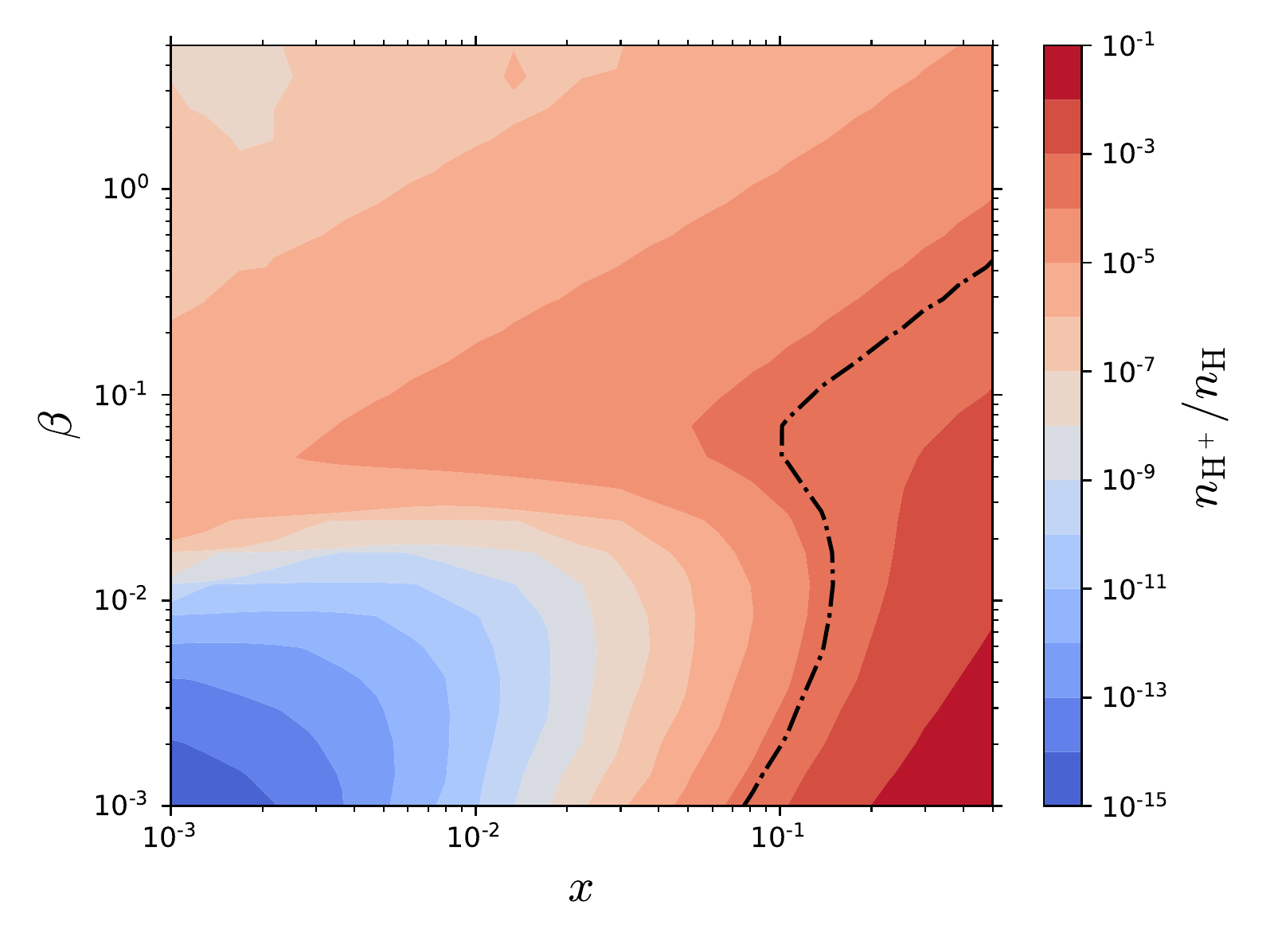}\hfil
		\includegraphics[width=0.5\linewidth]{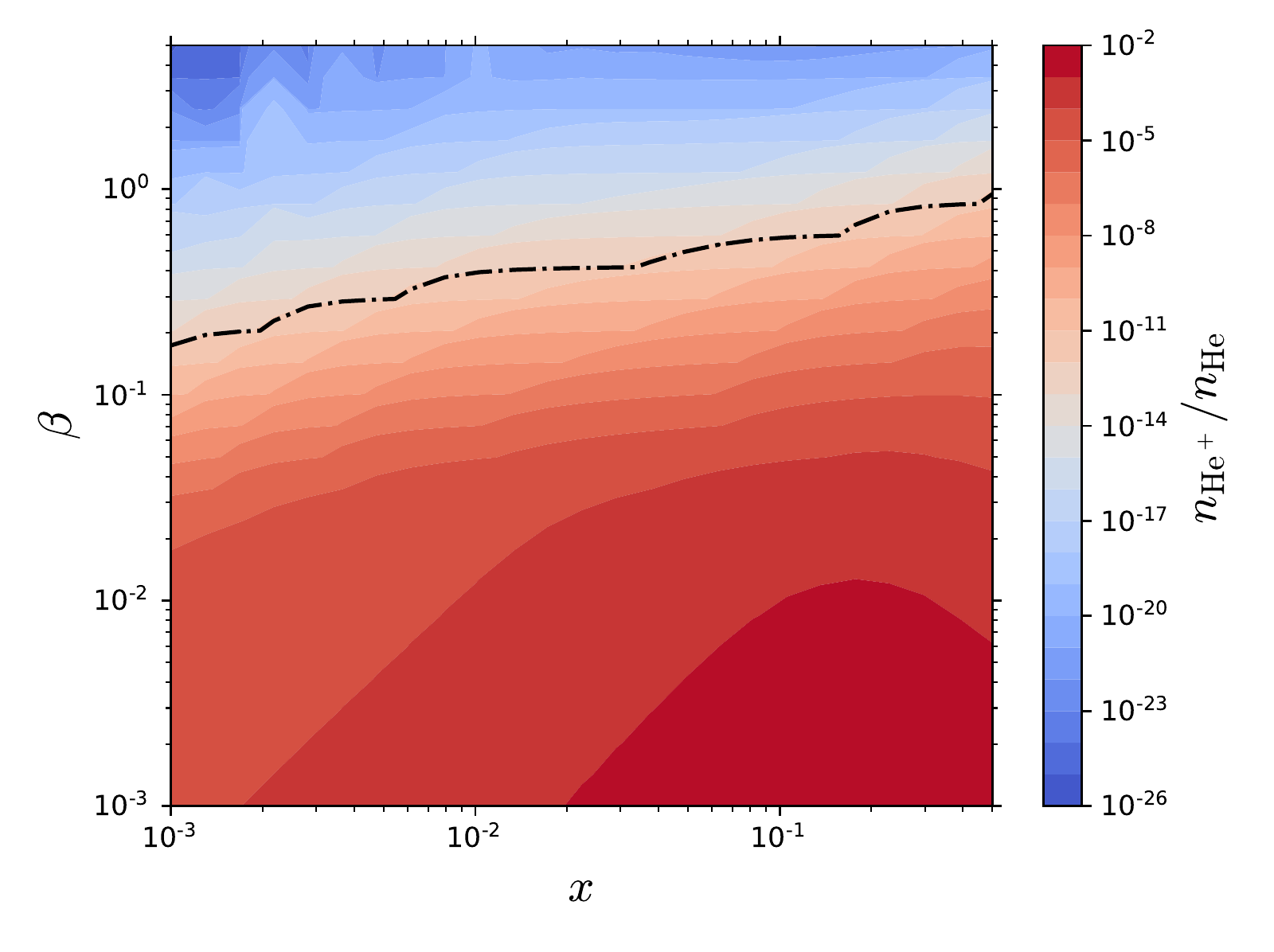}
		\caption{Residual ionization fractions $n\hii/n\h$ (left) and $n\heii/
		n\he$ (right) of the mirror sector at $z=10$. The dot-dashed 
		contours indicate the SM values: $n\hii/n\h= 2.2\times 10^{-4}$ and 
		$n\heii/n\he=1.2\times10^{-12}$.}
		\label{fig:recRes}
	\end{center} 
\end{figure*}

Fig.\ \ref{fig:recRes} shows the $(x,\beta)$-dependence of the 
residual ionization fractions of H and 
He at $z=10$. For comparison, the
SM values (at $x=1,\,\beta=1,\,Y=0.24$) are $n\hii/n\h= 2.2\times 10^{-4}$ 
and 
$n\heii/
n\he=1.2\times10^{-12}$.
Recombination of He is more efficient (blue regions) for high $\beta$, because
a larger mirror matter density increases the collision rate between ions and 
free electrons. As $x$ decreases, the interval between the beginning of 
recombination ($z_{\mathrm{rec}}'\simeq 1100/x$) and $z=10$ becomes longer, 
increasing the number of occasional ion-electron collisions following 
freeze-out. This and the slightly larger value of $n\he$ explain the 
somewhat higher efficiency of He recombination at low $x$.

We can also understand qualitative features of fig.\ \ref{fig:recRes}
concerning H recombination.  In contrast to He, there is a much stronger
variation of $n\h$, which changes by a factor of 60 as $x$ goes from $10^{-3}$ to
$0.5$, as compared to only a factor of 2 variation in $n\he$.
In particular, for $x\ll 1$ the low density of H is 
overwhelmed by free electrons, requiring relatively few 
collisions to recombine such that H may become neutral before 
He does so.
In this situation, for  $\beta\sim 1$, H recombination takes place much 
earlier than for He
and it is more efficient than in the SM. But since He recombines very 
effectively,  the number of free electrons available for
hydrogen-electron collisions after the freeze-out drops significantly, 
leading to a much higher ionization fraction than for He.
For $\beta\ll 1$, He recombination is very inefficient, leaving a larger number 
of free electrons to combine with H, and leading to a small ionization 
fraction. In the region where $x\gtrsim0.1$, hydrogen and helium number 
densities are almost equal, and their ionization fractions display a similar 
qualitative dependence on $\beta$.

\subsection{H$_2$ formation}

H$_2$ is an important molecular species for structure formation since it can
cool a primordial gas cloud to a temperature as low as $\sim 200$ K. Even in the
\MS, a small fraction of H$_2$ can act as an effective heat sink
that drives the collapse of large clouds into stellar objects. Conversely,
without  H$_2$, a virialized gas cloud of mirror helium might not cool below a 
temperature of order $10^4$ K (about 1 eV, or roughly 10~\% of ionization 
energy of helium and hydrogen), preventing structure formation.

\begin{figure}[b]
	\begin{center}
		\includegraphics[width=\linewidth]{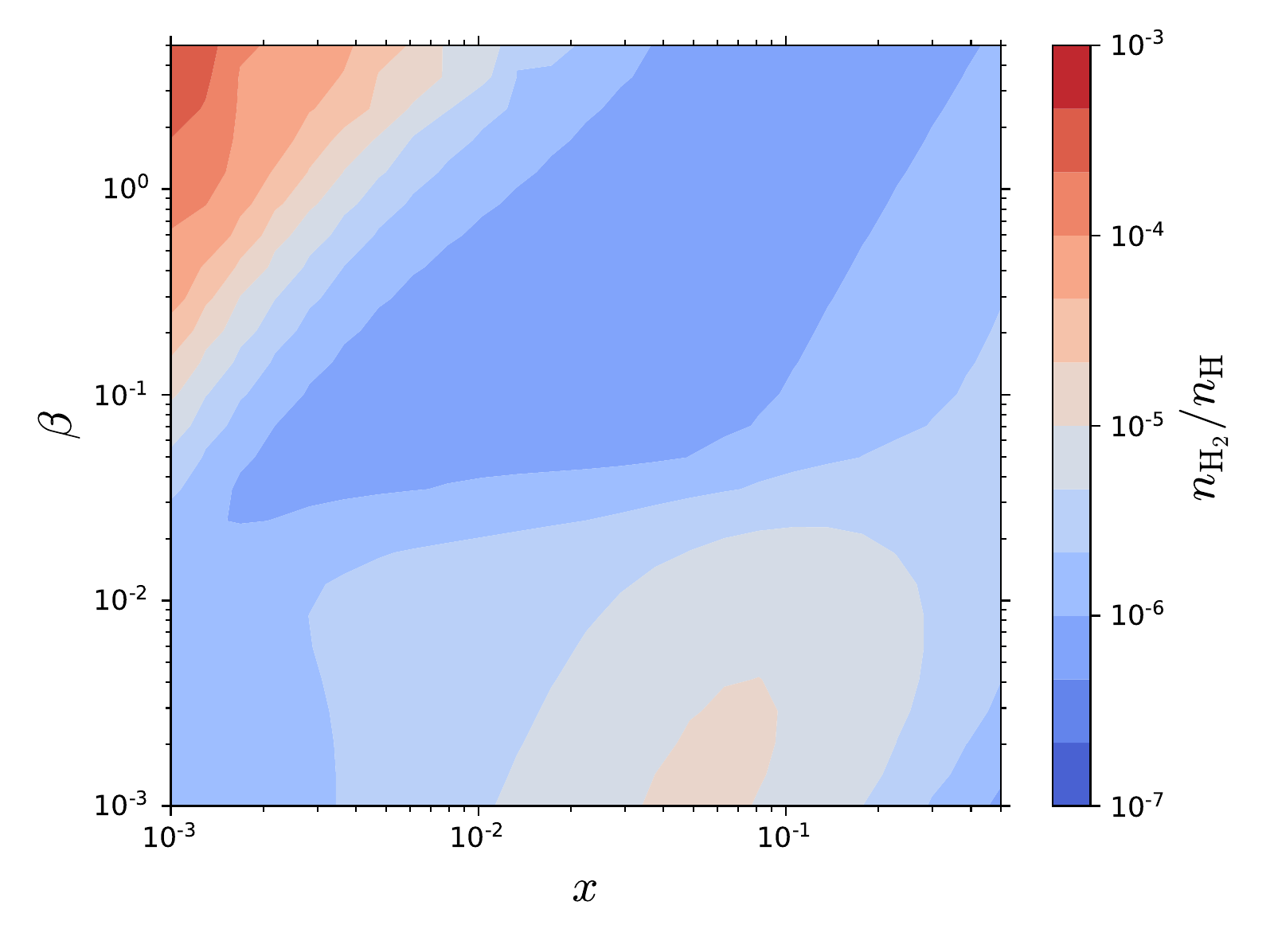}
		\cprotect\caption{Residual fraction $f_2=n\hh/n\h$ produced during 
		mirror 
		recombination at $z=10$. This fraction is higher than the SM value 
		of $n\hh/n\h \simeq 6 \times 10^{-7}$ for any ($x,\beta$).}
		\label{fig:recH2}
	\end{center} 
\end{figure}

Since H$_2$ has no dipole moment, it cannot form directly from the 
collision of two neutral H atoms. 
Instead, at early times its formation proceeds through
the reactions
\be 
\begin{gathered}\label{eq:h2form}
	\mathrm{H}^0 + e^- \rightarrow \mathrm{H}^- + \gamma, \\
	\mathrm{H}^- + \mathrm{H}^0 \rightarrow \mathrm{H}_2 + e^-.
\end{gathered}
\ee 
H$_2$ is always energetically favored at low temperatures, but the low matter 
density and ionization fractions inhibit its production after 
recombination. Hence H$_2$ can only form during 
recombination, when both $n_e$ and $n\hi$ are significant.

Other mechanisms involving H$_2^+$ and HeH$^+$ are known to contribute to  the
residual H$_2$ abundance, but these processes are subdominant 
\cite{Hirata:2006bt}. At late times, after the first generations of stars,
H$_2$ formation is catalyzed by dust grains and proceeds more rapidly,
but  between these epochs the reactions  (\ref{eq:h2form}) are the only
available route. 

As was shown in ref.\ \cite{Hirata:2006bt}, the production of H$_2$ depends on the 
abundance of H$^-$, which in the steady-state approximation is:
\be\label{eq:hmSteady}
X\hm = \frac{k_7 X_e X\hi n\h}{k_{-7} + k_8 X\hi n\h + k_9 X_e n\h+ k_{15} 
X\hii n\h}.
\ee
The rates $k_i$ are listed in 
table \ref{tab:rates}.

The residual H$_2$ abundance is determined by the Boltzmann equation
\be \label{eq:h2rec}
\frac{dX\hh}{dt} = k_8 X\hi X\hm n\h.
\ee
Since both H$^-$ and H$_2$ attain low abundances, their presence has 
little effect on the evolution of recombination. We can integrate 
eq.\ (\ref{eq:h2rec}) using the numerical method from the previous 
subsection. The fraction of $f_2' = n\hh/n\h$ produced by $z=10$ is illustrated in 
fig. \ref{fig:recH2}. For reference, the same analysis in the SM yields 
$f_2 \simeq 6 \times 10^{-7}$.
We find that $f_2'$  is always greater than $f_2$, 
analogously to the higher efficiency of mirror 
recombination. The degree of enhancement $f'_2/f_2$ depends on the timing of He 
recombination versus that of H, since H$_2$ requires both neutral H
and free $e^-$ for its formation.

When $\beta,x\sim 1$, recombination proceeds similarly as in the  
SM: He recombines efficiently and prior to H, leaving too few
$e^-$ for H$_2$ to form. As $\beta$ decreases, He recombination 
becomes incomplete and the extra $e^-$ density produces more H$_2$. 
For $x\ll 1$ but $\beta\sim 1$, H recombines before He, leading
to simultaneously high abundances of neutral H and free $e^-$. This 
explains the enhanced H$_2$ production in fig.\ \ref{fig:recH2}. If both $x\ll1$ 
and $\beta\ll1$, the two recombinations overlap, leaving fewer $e^-$ 
to produce molecules.

\section{Mirror matter structure formation}
\label{sec:sf}

We use the semi-analytical galaxy formation model GALFORM presented in 
ref.\ 
\cite{Cole:2000ex} to predict structures in the \MS. Our analysis
parallels that of ref.\ \cite{Ghalsasi:2017jna}, but is complicated by the
additional chemical elements present in the \MS\ relative to the simple
atomic DM model considered there.
In particular, nuclear reactions in the \MS\ allow for the formation of mirror 
stars and 
supernovae whose feedback can impact the collapse of gas clouds.

\subsection{Merger tree}
Our current understanding of structure formation is that galaxies 
formed following a bottom-up hierarchy: small halos merged at early times and 
grew into larger overdense regions. The extended Press-Schechter formalism 
\cite{Press:1973iz,Bond:1990iw,Lacey:1993iv}, which we summarize here, is an 
analytic description of the statistical growth and merger history of a halo that 
reproduces the results of cosmological simulations.

Let $M_2$ be the mass of a halo at time $t_2$. The mass fraction $f_{12}
(M_1,M_2) \, dM_1$ of $M_2$ that was in halos in the interval $\left[M_1, M_1 
+ dM_1\right]$ at a time $t_1<t_2$ is
\begin{align}
\label{eq:f12}
f_{12}\left(M_{1}, M_{2}\right) d M_{1}&=\frac{1}{\sqrt{2 \pi}} \frac{\left(\delta_{\mathrm{c} 1}-\delta_{\mathrm{c} 2}\right)}{\left(\sigma_{1}^{2}-\sigma_{2}^{2}\right)^{3 / 2}}\\
&\times \exp \left(-\frac{\left(\delta_{\mathrm{c} 1}-\delta_{\mathrm{c} 2}
\right)^{2}}{2\left(\sigma_{1}^{2}-\sigma_{2}^{2}\right)}\right) 
\frac{d \sigma_{1}^{2}}{d M_{1}} d M_{1}, \nn
\end{align}
where $\sigma^2(M)$ is the variance of the matter power spectrum $P(k)$
inside a sphere of comoving radius $R= (3M/4\pi \rho_m)^{1/3}$, extrapolated 
linearly to $z=0$, and $\delta_{\mathrm{c}}(t)$ is the critical overdensity for 
gravitational collapse at time $t$, also extrapolated to current times,
\be \label{eq:deltac}
\delta_{\mathrm{c}}\left(t_{\mathrm{c}}\right)=\frac{1}{D(z_\mathrm{c})}
\frac{3}{5}\left(\frac{3 \pi}{2}\right)^{2 / 3}
\left[\Omega_{\mathrm{m}}\left(z_{\mathrm{c}}\right)\right]^{0.0055}.
\ee 
The linear growth factor $D(z)$ (set to unity at $z=0$) evolves as
\be 
D(z) \propto H(z) \int_{z}^{\infty} \frac{1+z'}{H(z')^3}\, dz'
\ee
and $z_c$ is the redshift at time $t_c$. In a matter-dominated 
universe $D(z)$ is exactly equal to the scale factor $a=(z+1)^{-1}$, 
but here we
also account for dark energy, which becomes important as
$z\to 0$.

Taking $t_2 = t_1 + dt$ with $dt$ arbitrarily small, eq.\ 
(\ref{eq:f12}) becomes
\be  \label{eq:df12}
\frac{d f_{12}}{d t} =\frac{1}{\sqrt{2 \pi}} 
\frac{1}{\left(\sigma_{1}^{2}-\sigma_{2}^{2}\right)^{3 / 2}} 
\frac{d \delta_{\mathrm{c} 1}}{d t} \frac{d \sigma_{1}^{2}}{d M_{1}}.
\ee 
Therefore the average number of objects in $\left[M_1, M_1 + 
dM_1\right]$ that combined during $dt$ to form the 
halo of larger mass $M_2$ is
\be \label{eq:dn}
dN = \frac{df_{12}}{dt} \frac{M_2}{M_1} \, dt \, dM_1.
\ee
The algorithm presented in refs.\ \cite{Cole:2000ex,Ghalsasi:2017jna} uses 
eq.\ 
(\ref{eq:dn}) to find the progenitors of a halo of mass $M_2$ by taking small 
steps $dt$ backwards in time. The resulting ``merger tree'' describes the 
hierarchical formation of the halos observed at $z=0$. 

Numerically, one must define a resolution scale $M_{\mathrm{res}}$ below 
which there is no further tracking of individual halos. The probability that a halo of mass 
$M_2$ splits into halos of masses $M_1 \in \left[M_{\mathrm{res}}, M_2/2
\right]$ and $(M_2-M_1)$ in a backward step $dt$ is
\be \label{eq:Pmerger}
P = \int_{M_{\mathrm{res}}}^{M_2/2} \frac{dN}{dM_1}\, dM_1.
\ee 
Accretion of objects smaller than $M_{\mathrm{res}}$ also contribute to the 
growth of the halo during that period. The fraction of mass that is lost to 
those smaller fragments in the reverse time evolution is
\be \label{eq:Fmerger}
F = \int_{0}^{M_{\mathrm{res}}} \frac{dN}{dM_1} \frac{M_1}{M_2}\, dM_1.
\ee 

The algorithm to generate the merger tree is as follows. Starting at redshift $z_f$
with a single halo of mass $M_2=M_f$, a backward time step $dt$ is taken, with
$dt$ small enough that $P\ll1$. A random number $R$ is generated from a uniform 
distribution between $0$ and $1$. If $R>P$, the halo does not fragment, but still 
loses a fraction of mass $F$ due to the accretion of matter below the resolution
scale. Thus the mass of the halo at the next time step becomes $(1-F)M_2$. If 
$R<P$, the halo splits into two halos of mass $M_1$ and $(1-F)M_2 -M_1$ where 
$M_1$ is chosen randomly from the distribution given by eq.\ \ref{eq:dn}. These
steps are repeated for every progenitor whose mass is above $M_{\rm res}$
until the chosen initial redshift $z_i$ is reached.

We used this algorithm to generate 10 merger trees for a final halo mass of
$M_f=10^{12}\ M_\odot$, about the size of the Milky Way.  The
time interval between $z_f=0$ and $z_i=10$ was divided into $10^4$ 
logarithmically scaled time steps. The resolution was set to 
$M_{\mathrm{res}} =3\times 10^7\ M_\odot$, well below $M_f$ but large 
enough to
avoid keeping track of too many halos simultaneously. To minimize possibly
large statistical fluctuations, we used the ensemble of merger trees to 
average over all derived quantities in the end.  Inspection of the individual
trees indicated that 10 was more than sufficient to avoid spurious effects
of outliers. 

Neither the distribution nor the nature of matter inside the halos affects
the evolution of the merger tree. Therefore the algorithm described above is 
completely model-independent, to the extent that matter overdensities are 
Gaussian. This allows us to use the same 10 merger trees in scanning over all 
values of $x$ and $\beta$ for structure formation. However $P(k)$
depends on the nature of dark matter, which in turn
affects the variance $\sigma^2(M)$ in eqs.\ 
(\ref{eq:f12},\ref{eq:df12}-\ref{eq:Fmerger}). 
For simplicity, we computed $\sigma^2$ with 
\verb|Colossus| \cite{Diemer:2017bwl}, but this package assumes a $\Lambda$CDM 
cosmology. For self-consistency, it is necessary to verify that
$P(k)$ and its variance $\sigma^2$ do not differ too much from their standard cosmology
expressions in the presence of a mirror sector. We discuss this issue below.

\subsection{Mirror Silk damping}
\label{sec:Silk}

In the early universe, photons and baryons are tightly coupled, making the
mean free path of photons $\lambda_\gamma$ negligible, but at the onset of
recombination $\lambda_\gamma$ becomes significant. Photons can then
diffuse out of overdense regions, effectively damping perturbations on scales smaller 
than the Silk scale $\lambda_D$, which we derive below. In the SM, the mass 
scale corresponding to the Silk length is 
$M_D\sim10^{12}\ M_\odot$ \cite{Kolb:1990vq}, about the mass of the 
Milky Way halo. Structure formation below this scale is 
strongly inhibited, unless a significant component of CDM allows small-scale 
perturbations to grow.

Mirror matter can be similarly affected by collisional damping. Since 
we observe structures on scales smaller than $M_D$, Silk damping sets a 
lower bound on the amount of ordinary CDM required for the mirror model to agree with current 
data. A full analysis of cosmological perturbations, acoustic oscillations and 
the matter power spectrum is outside the scope of this work. However, the 
equations presented in the previous section depend on $P(k)$ through
its variance $\sigma^2(M)$. We must therefore  
check that $P(k)$  is not too 
different from its $\Lambda$CDM value. Many effects 
could alter $P(k)$, like extra oscillations on scales smaller than the sound 
horizon of the mirror matter plasma \cite{Cyr-Racine:2013fsa}, but Silk damping 
has the largest impact on our structure formation analysis.
In particular, small-scale perturbations must be able to grow 
sufficiently for galaxy formation to proceed hierarchically.  Hence we estimate 
the size of the \MS\ counterpart of the damping scale, $\lambda'_D$, and its 
implications for the growth of \MS\ density perturbations. 

\subsubsection{Mirror Silk scale}

One can estimate the \MS\ Silk scale as follows \cite{Kolb:1990vq}. The mean 
free path
of \MS\ photons at low temperatures is
\be \label{eq:photonmfp}
 \lambda_{\gamma'} = \frac{1}{n_e \sigma_T} = \frac{1}{\xi_e n_N \sigma_T},
\ee
where $\xi_e\equiv n_e/n_N$ is the ionization fraction during H and He$^+\to$ 
He$^0$ 
recombination. During an interval 
$\Delta t$, a photon experiences $N=\Delta t/\lambda_{\gamma'}$ collisions. 
The average comoving distance $\Delta r$ traveled in this time is that  of a random walk with a 
characteristic step of length $\lambda_{\gamma'}/a$,
\be 
(\Delta r)^2 = N \frac{\lambda_{\gamma'}^2}{a(t)^2} = 
\frac{ \lambda_{\gamma'} \Delta t}{a(t)^2}.
\ee 

Taking the limit $\Delta t\rightarrow0$ and integrating until recombination gives
\begin{align} \label{eq:silk}
\begin{split}
\lambda_D^{\prime2} &= \int_{0}^{t_{\mathrm{rec}}} \frac{\lambda_{\gamma'}}{a(t)^2} dt \\
&\simeq - \lambda_{\gamma'}(z'_{\mathrm{rec}})\ 
\left(1+z'_{\mathrm{rec}}\right)^3 
\int^{\infty}_{z_{\mathrm{rec}}'} \frac{1}{1+z} \left(\frac{dt}{dz}\right) dz,
\end{split}
\end{align}
using the fact that $\lambda_{\gamma'}$ scales as $n_N^{-1}\sim a^3$ and 
approximating $\xi_e$ as constant during the period where  $\lambda_{\gamma'}$ is 
large.

Recalling that the redshift of mirror recombination is 
$z_{\mathrm{rec}}' \simeq 1100/x$, this occurs before matter-radiation equality 
($z_{\mathrm{eq}}=3365$)\footnote{Our bound on $x$ from CMB and BBN ensures 
that $z_{\mathrm{eq}}$ doesn't change significantly due to the presence of 
mirror radiation.} for $x\lesssim 0.3$, and during the early matter-dominated era 
otherwise. For simplicity, consider the case $x\ll0.3$ so that 
mirror recombination 
completes during  the radiation-dominated era when $t\sim(1+z)^{-2}$. Eq. 
(\ref{eq:silk}) then reduces to
\be  \label{eq:silkApprox}
\lambda_D^{\prime 2} \simeq \frac{2}{3} t_{\mathrm{rec}}'\ \lambda_{\gamma'}
(z_{\mathrm{rec}}')\ (1+z_\mathrm{rec}')^2, \ \quad x\ll0.3.
\ee 
In the case where recombination occurs much later than 
$z_{\mathrm{eq}}$ (like in the SM), we would obtain the same expression, without
the primes, up to 
the numerical coefficient \cite{Kolb:1990vq}. This implies that $\lambda'_D\ll 
\lambda_D$, unless $\beta$ is very small and the mirror matter plasma is 
diluted before recombination.

To further quantify $\lambda'_D$, we note that at early times when vacuum 
energy is
negligible so that $\Omega_m+\Omega_{\mathrm{rad}}=1$, $t(z)$ is given by

\begin{align} \label{eq:timeMDRD}
\begin{split}
t(z) &= \frac{2}{3 H_0 \sqrt{\Omega_{\mathrm{m,0}}}} \frac{1}{(1+z_{\mathrm{eq}})^{3/2}} \\
&\times
\left[2 + \left(\frac{1+z_{\mathrm{eq}}}{1+z}-2\right)\sqrt{\frac{1+z_{\mathrm{eq}}}{1+z}+1}\right].
\end{split}
\end{align} 
which for $z\gg z_{\mathrm{eq}}$ simplifies to
\be 
t(z\gg z_{\mathrm{eq}}) \simeq \frac{1}{2 H_0 \sqrt{\Omega_{\mathrm{m,0}}}}  
\frac{\sqrt{1+z_{\mathrm{eq}}}}{(1+z)^2}.
\ee
Then eq.\ (\ref{eq:silkApprox}) can be rewritten in terms of 
$x (\ll 1)$ and $\beta$ as
\be \label{eq:silkApp2}
\lambda_D^{\prime 2} \simeq \frac{8 \pi G}{9 H_0^3 \sqrt{\Omega_m} \Omega_b} 
\left(\frac{\mn x^3}{\xi_e \sigma_T \beta }\right) 
\frac{\sqrt{1+z_{\mathrm{eq}}}}{(1+z_{\mathrm{rec}})^2}.
\ee 
(recall eq.\ (\ref{eq:mn}) for $\mn$), where we used $\xi_e\sim0.1$ 
at the time of recombination. Hence $\lambda'_D$ scales as $(x^{3}/\beta)^{1/2}$.
For larger values of $x$, $z_{\mathrm{rec}}'$ is close to $z_{\mathrm{eq}}$ and 
we cannot assume a fully matter- or radiation-dominated universe to compute the 
integral of eq.\ (\ref{eq:silk}); nevertheless we verified that eqs.\ 
(\ref{eq:silkApprox},\ref{eq:timeMDRD}) are accurate to within several percent
even for $x>0.3$. 

\subsubsection{Growth of MS perturbations}

The previous estimate for $\lambda'_D$ allows us to predict the growth of
density perturbations in the \MS.
Consider a mirror baryonic overdensity $\delta_{b'}(k) = (\rho_{b'}(k) - 
\overline{\rho_{b'}} ) /\overline{\rho_{b'}}$ on a scale $\lambda = \pi/ k$. 
Assuming primordial perturbations are adiabatic, we have $\delta_{b'} =
\delta\cdm$ at early times ($z\gg z_{\mathrm{eq}}$). $\delta\cdm$ remains
nearly constant prior to matter-radiation equality (ignoring small 
logarithmic growth of subhorizon modes). However Silk damping 
suppresses $\delta_{b'}(k)$ by a factor $\sim \exp \left(-k^2/k_D'^2\right)$ 
after recombination \cite{MBW:2010}, where $k_D'=\pi/\lambda_D'$.

For very small values of $\beta$, the \MS\ constitutes only a small fraction 
of DM and the power spectrum is not significantly affected by mirror Silk 
damping. Therefore in what follows we only consider $\beta \gtrsim 0.1$, where 
consequently $\lambda_D \gg \lambda'_D$. The analysis below focuses on scales 
sufficiently small so that SM baryonic perturbations are always negligible 
compared to CDM and \MS\ overdensities.

Starting at $z=z_{\mathrm{eq}}$, both the CDM and mirror components grow 
linearly according to
the cosmological perturbation equations \cite{Naoz:2005pd}
\begin{gather} \label{eq:perturbCDM}
	\ddot{\delta}\cdm + 2 H \dot{\delta}\cdm = 
\frac{3}{2} H_0^2 \frac{\Omega_{\mathrm{DM}}}{a^3} \left(f_{b'} \delta_{b'} + 
f\cdm \delta\cdm\right)\\
	\ddot{\delta}_{b'} + 2 H \dot{\delta}_{b'} = \frac{3}{2} H_0^2 \frac{\Omega_{\mathrm{m}}}{a^3} \left(f_{b'} \delta_{b'} + f\cdm \delta\cdm\right) - \frac{k^2}{a^2} c_s^2 \delta_{b'},  \label{eq:perturbBar}
\end{gather}
where $f_i\equiv \Omega_i/\Omega_{\mathrm{DM}}$ are the fractions of the 
total DM density, such that $f_{b'} + f_c = 1$. We omit
the equation  for $\delta_b$, which is highly damped on small scales.

Recall that mirror baryons recombine before $z_\mathrm{eq}$ for 
$x\lesssim0.3$. Subsequently the \MS\ pressure drops 
precipitously,  which means that its sound speed $c'_s\sim 0$ at matter-radiation equality. 
Even before \MS\ recombination, $c_s^{\prime 2}$ is suppressed 
by a factor $\sim x^4$ compared to the SM $c_s^2$, due to the low \MS\ 
temperature 
\cite{Berezhiani:2000gw,Ignatiev:2003js}. To a first approximation we can 
therefore ignore the pressure term in eq.\ (\ref{eq:perturbBar}) for all values of 
$x$. We can then combine the two ODEs into
\be 
\ddot{\delta}'+ 2 H \dot{\delta}' = \frac{3}{2} H_0^2  
\frac{\Omega_{\mathrm{DM}}}{a^3} \delta'
\ee 
where $\delta' = (f_{b'}\delta_{b'} + f\cdm \delta\cdm)$ is the total matter 
perturbation; the CDM and mirror matter perturbations 
evolve together during the matter-dominated era and their ratio remains
constant. The growing mode grows as $\delta'\sim a$, so that at small 
redshifts
\begin{align}\begin{split}
\delta'(z) =& \delta'(z_{\mathrm{eq}}) \left(\frac{1+z_{\mathrm{eq}}}{1+z}\right)\\
 =& \delta\cdm (z_{\mathrm{eq}})\, \left(1-f_{b'} (1-e^{-k^2/k_D'^2})\right) 
 \left(\frac{1+z_{\mathrm{eq}}}{1+z}\right),
 \end{split}
\end{align}
where we combined the initial abadiatic condition $\delta_{b'} =\delta\cdm$ 
with the exponential Silk damping. Therefore we see that Silk damping 
suppresses small scale matter perturbations by an additional factor of roughly
\be 
\mathcal{F}_D = \left(1-\frac{\beta \ob}{\Omega_{\mathrm{DM}}} 
(1-e^{-k^2/k_D'^2})\right) 
\ee
compared to standard cosmology.

\subsubsection{Effect on the merger tree evolution}

To verify that our merger tree evolution is not significantly altered by 
the 
suppression of $P(k)$ on small scales, we applied the Silk damping factor 
$\mathcal{F}_D$ to the $\Lambda$CDM matter 
power spectrum and we computed the variance 
$\sigma^2(M)$ and the integral $P$ of eq.\ (\ref{eq:Pmerger}) 
for a Milky Way-like halo ($M_2=10^{12}$ M$_\odot$) with this extra feature. 

The value of $P$ is the probability for a merger to happen and it is roughly 
inversely proportional to the lifetime of large halos $t_{\mathrm{halo}}$, 
which we will properly define later. The accretion rate $F$ given by eq.\ 
(\ref{eq:Fmerger}) also affects $t_{\mathrm{halo}}$, but for large halos it 
represents such a small fraction of the total mass that we can ignore it. Let 
$P_D$ be the value of the integral of eq.\ (\ref{eq:Pmerger}) computed with the 
damped power spectrum.  We expect that the lifetime should scale 
  as $t_{\mathrm{halo}}\sim P/P_D$. 

In our 10 merger trees, the average lifetime of the Milky Way halo is 
6.9 Gyr with a relative standard deviation of 21.5~\%. To ensure the 
self-consistency of our analysis, we demand that the Silk damping does not 
change the average lifetime by more than $2\sigma$, or 43~\%. In other words, 
our analysis is valid only if $0.57 < P/P_D < 1.43$; outside this region we 
cannot trust our conclusions because the merger trees would be too drastically 
affected by the damping effects.

We find that two regions above $\beta\gtrsim3.7$ must be excluded from our 
analysis: for $0.02\lesssim x\lesssim 0.12$,  $t_{\mathrm{halo}}$ 
would be much longer than the estimate we obtained using the $\Lambda$CDM power 
spectrum; whereas for $x\gtrsim 0.2$, the halo lifetime in the 
presence of mirror matter would be too small. These regions are illustrated 
with our results on fig.\ \ref{fig:results}.

Note that the different behaviors in these two regions come from two 
competing effects in eq.\ (\ref{eq:df12}): both $\left|d\sigma_1^2/dM_1\right|$ 
and $(\sigma_{1}^2-\sigma^2)^{3/2}$ are suppressed by the collisional damping, 
but the latter effect dominates for large values of $x$, when the Silk scale is 
large. Interestingly, those effects cancel out around $x\simeq0.15$ and we can 
still use our structure formation analysis to constrain the scenario where 
mirror matter makes up all DM in this region. Fortunately, the high temperature 
region also corresponds to the parameter space that is more likely to be 
constrained by cosmological observables like the CMB or the matter power 
spectrum. Ref. \cite{Ciarcelluti:2010zz} already constrained $x\lesssim 0.3$ if 
mirror matter were to make up all DM.

\subsection{Virialization and cooling}
\label{virial} 
Once linear matter perturbations exceed the critical overdensity 
$\delta_{\mathrm{c}}$ (eq.\ (\ref{eq:deltac})) they collapse gravitationally 
into a virialized halo whose average overdensity is \cite{MBW:2010}: 
\be
\Delta\vir(z) = \frac{18\pi^2+82\, y-39\, y^2}{\Omega_{\mathrm{m}}(z)},
\ee 
with $y = \Omega_{\mathrm{m}}(z)-1$. Since mirror baryonic matter 
is not pressureless, this collapse leads to accretion shocks that heat the gas 
to a temperature of roughly \footnote{The expression for $T\vir$ is for a 
truncated isothermal halo, which requires an effective external pressure term 
to be in equilibrium. Without external pressure the numerical coefficient would 
be 1/5 instead of 1/2, which might be more familiar to the reader.}
\footnote{We omit primes in this section, where the formalism applies equally to
SM or \MS\ baryons.} 
\cite{MBW:2010,DAmico:2017lqj}:
\be \label{eq:virTemp}
T_M = (\gamma-1 )\ T\vir = (\gamma-1)\ \frac{1}{2} \frac{G M\mu}{r\vir},
\ee  
where $\gamma$ is the adiabatic index of the gas, $\mu$ is the mean 
molecular mass, $M$ is the total mass of the halo (including the CDM component) 
and $r\vir$ is the virial radius. We define $r\vir$ as the radius of a sphere 
inside which the average overdensity is equal to $\Delta\vir$. For simplicity, 
we will assume the gas is purely monatomic, which sets $\gamma = 5/3$.

The temperature of a virial halo is always much greater than the temperature of 
the matter background. This means the baryonic pressure $P=\rho T/\mu$ becomes 
nonnegligible and prevents further collapse of mirror matter. In order for 
galaxies to form, mirror baryons must radiate energy, which is why structure 
formation is impossible without an efficient cooling mechanism. 

Let $\cool_i = - du/dt$ be the cooling rate of a given process $i$, where $u$ is the 
energy density of the gas ($\cool_i$ is positive if the energy is lost). If 
several reactions contribute to the total $\cool$, we can define the cooling timescale 
as
\be \label{eq:tcool}
t_{\mathrm{cool}}(r) = \frac{3}{2} \frac{n(r)T_M}{\sum_i \cool_i(r,T_M)}.
\ee 
where $n(r)$ is the number density of all chemical species combined. 
Therefore $t_{\mathrm{cool}}$ is 
roughly the time required for the gas to radiate all its kinetic energy. 
Since the gas is not homogeneous, the cooling timescale decreases as we move 
further away from the center of the halo.
We describe
the various contributions to $\cool$ 
\cite{MBW:2010,Cen:1992zk,Grassi:2013lha,1979ApJS...41..555H} in 
appendix \ref{appB}.

To compute the cooling rates and timescale, one must also specify 
the number density $n_i$ of each chemical species. In general, their relative 
abundances are determined by rate equations of the form 
\cite{Grassi:2013lha}
\be \label{eq:rateEq}
\frac{dn_i}{dt} = \sum_{j\in F_i}\left(k_j \prod_{r\in R_j} n_r^{(j)}\right) - 
\sum_{j\in D_i}\left(k_j \prod_{r\in R_j} n_r^{(j)}\right),
\ee 
where $F_i$ and $D_i$ are the sets of reactions $R_j$ that form and destroy the 
$i$th species and $n_r^{j}$ is the number density of each reactant in $R_j$. 
The coefficients $k_j$ set the rate of each reaction and usually depend on the 
temperature of the system. If the right-hand-side of eq.\ (\ref{eq:rateEq}) vanishes for a 
given species, the reaction is in collisional equilibrium, or steady 
state. If all processes are two-body reactions, the steady-state 
density is given by
\be \label{eq:steady}
n_i = \frac{ \sum\limits_{j\in F_i}k_j n_{1}^{(j)} n_{2}^{(j)}}{ 
\sum\limits_{j\in 
D_i} k_j n_d^{(j)}}.
\ee  

The cooling mechanisms depend on the abundances of eight chemical species: H$^0$, 
H$^+$, H$^-$, H$_2$, He$^0$, He$^+$, He$^{++}$ and $e^-$ \footnote{Reaction 11 in 
table \ref{tab:rates} produces H$_2^+$ but we did not consider any cooling 
mechanism associated with this ion. Since its abundance is negligible at all 
times we omit it from our analysis.}. In 
the steady-state approximation, the network eq.\ (\ref{eq:steady})  
is usually underdetermined, but one can solve it
if 1) the total nuclear density 
$n_N=n\h+n\he$ satisfies eq.\ (\ref{eq:nb}); 2) the total He-H number 
ratio $X\he=n\he/n\h$ satisfies eq.\ (\ref{eq:xhe}) (assuming nuclear reactions 
in stars do not strongly affect $X\he$); 3) matter is neutral, which implies 
$n_e \simeq n\hii + n\heii + 2 n\heiii$ (since the density 
of H$^-$ is negligible). The reactions considered in our simplified chemical 
network and their rates are given in table \ref{tab:rates}.

However, the steady-state approximation tends to break down
at low temperature/densities; if the timescale 
of a given reaction $t_{r}\sim (k_{r} n)^{-1}$ is smaller than the 
dynamical timescale of the system 
$t_{\mathrm{dyn}}\sim (G\rho)^{-1/2}$, the chemical species cannot reach 
collisional equilibrium. This is most likely to occur at early times 
in small halos with low density and temperature. At $z=10$ the
$t_{\mathrm{dyn}}$ corresponding to the virial overdensity is 
$\sim 0.2$ Gyr. Taking $n\sim 1$~cm$^{-3}$ --- roughly the 
central density in a $3\times10^7$ M$_\odot$ halo at this epoch --- we can 
check that H$_2$, H$^+$ and He$^+$ respectively come into 
collisional equilibrium at about 4,000~K, 9,000~K and 15,000~K. 
Below those critical temperatures we take the abundances of each species to be 
their relic densities after recombination, as determined by \verb|Recfast++|
(section \ref{sect:recomb}).

\begin{figure}[t]
	\begin{center}
		\includegraphics[width=\linewidth]{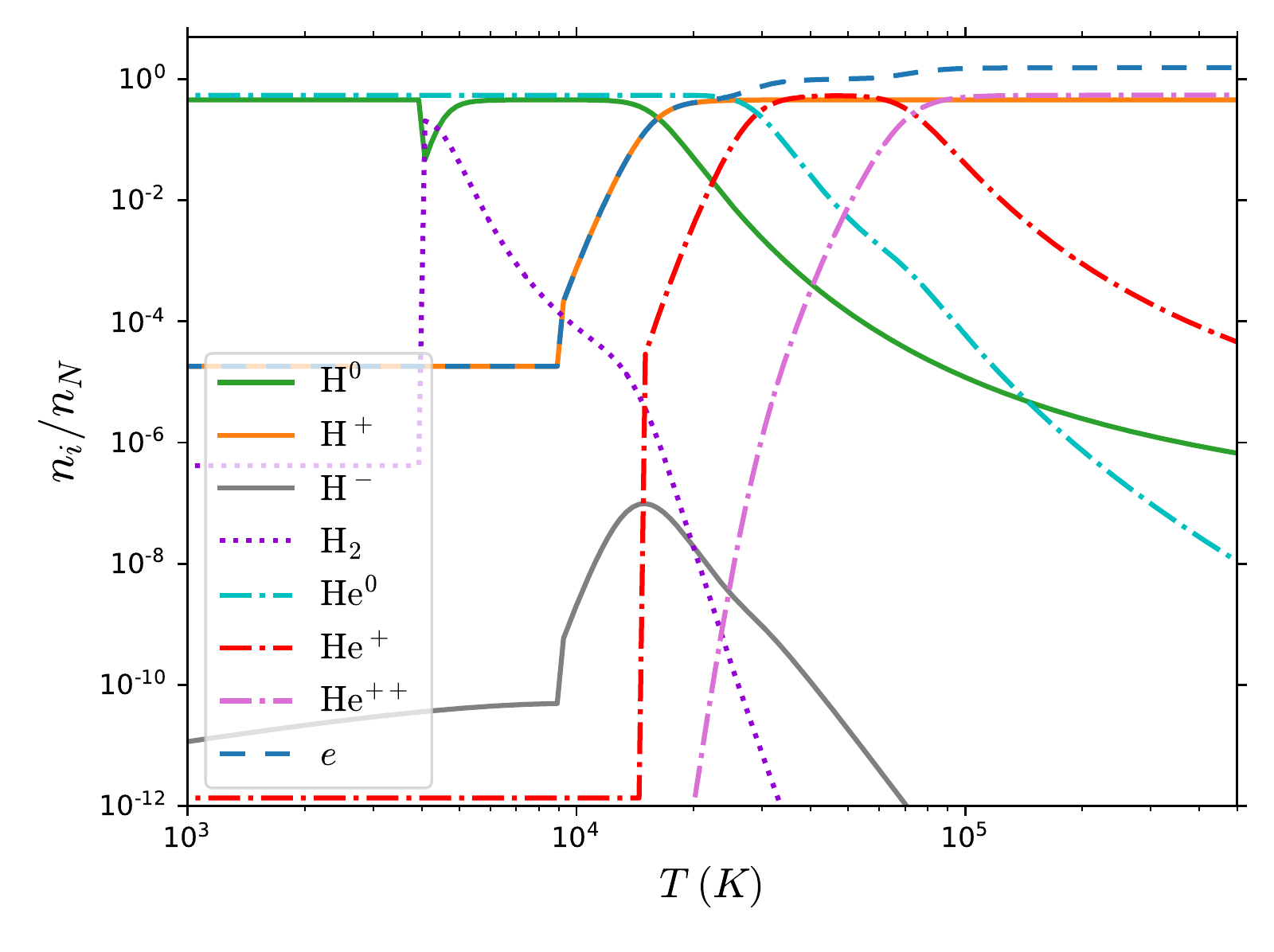}
		\caption{ Example of the relative 
			abundance of each chemical species at $z=10$ as a function of the 
			temperature of the gas. We used the benchmark parameters $x=0.1$ 
			and 
			$\beta=1$. }
		\label{fig:species}
	\end{center} 
\end{figure}

In reality the chemical species evolve toward their equilibrium values 
during shock heating; the true densities therefore lie between 
the equilibrium and the freeze-out values, but this difference has a negligible
impact on the cooling rates at high $z$, as well as on the
overall evolution of the galaxy.\footnote{He$^{++}$, which 
comes into equilibrium at $\sim$ 37,000~K, is a special case since we do not
solve for its relic density at recombination. Instead we take its
steady-state value at all temperatures, which has no 
effct on the cooling rates since its
abundance is negligible below 50\,000 K. We do likewise for 
H$^-$  since its high destruction rate keeps its abundance 
small at all times \cite{Hirata:2006bt}.}\ \  Fig.\ \ref{fig:species} 
illustrates the evolution of each chemical species with the 
temperature at $z=10$ with parameter $x=0.1$ and $\beta =1$.  The abrupt
transitions result from the approximations described here, and would be
smoothed out by fully solving eqs.\ (\ref{eq:rateEq}), but with no appreciable
effect on the consequent formation of structure.

\subsection{Cloud collapse and star formation}
\label{sec:starform}

As the cloud of mirror matter that fills the halo loses energy, its 
pressure drops and it is no longer sufficient to counteract the self-gravity of 
the gas. The cloud starts collapsing and fragments in overdense regions. If the 
cooling mechanism is very efficient, the collapse will 
occur on a characteristic timescale set by the free-fall time,
\be \label{eq:tff}
t_{\mathrm{ff}} (r) = \sqrt{\frac{3 \pi}{32 G \rho_r}}.
\ee 
In this expression $\rho_r$ is the average matter density inside a sphere of 
radius $r$. 

As the gas gets denser, the horizon of sound waves becomes increasingly 
smaller and matter cannot remains isothermal on scale larger than the Jeans 
length \cite{Ghalsasi:2017jna,CarrollOstlie},
\be
\lambda_J = \sqrt{\frac{15 k T}{4 \pi G \mu \rho}}.
\ee 
Below this scale the gas cannot fragment further. This sets the minimal 
mass of fragments that result from the collapse:
\be \label{eq:jeansmass}
M_J = \frac{4 \pi \rho}{3} \lambda_J^3.
\ee 

By evaluating the Jeans mass at the final density and temperature we can obtain 
a rough estimate of the mass of the primordial stars of mirror matter. 
Following 
\cite{DAmico:2017lqj}, we used \verb|Krome| \cite{Grassi:2013lha} to study the 
evolution of the temperature and the density of a collapsing cloud of mirror 
matter gas. \verb|Krome| assumes the cloud is in a free fall,
\be 
	{\dot{n}\over n}\sim {1\over t_{\mathrm{ff}}}
\label{ffeq}
\ee
(recall eq.\ (\ref{eq:tff}) for  $t_{\mathrm{ff}}$), 
and solves the 
out-of-equilibrium rate 
equations (\ref{eq:rateEq}). The temperature evolves as \cite{Grassi:2013lha}
\be 
\frac{\dot{T}}{T} = (\gamma-1) \left(\frac{\dot{n}}{n} + 
\frac{\mathcal{H}-\cool}{nkT}\right),
\label{Tevol}
\ee 
where  the 
cooling rate is $\cool = \sum_i\cool_i$ and  $\mathcal{H}$ is the heating 
rate. The heating rate is negligible in the optically thin limit because 
photons exit the cloud, but as the gas becomes denser we must include it in 
our calculations.  Recall that $\gamma=5/3$.

In the absence of cooling, eq.\ (\ref{Tevol}) shows that the
temperature of the cloud will increase as it collapses. However if the
cooling processes dominate, the collapse continues unimpeded as $T$
(and pressure) decreases. Eventually the gas becomes optically thick
and the cooling becomes ineffective; at this point the approximation
(\ref{ffeq}) of free-fall evolution breaks down and  the cloud can
only collapse adiabatically, which tends to increase the Jeans  mass.
In reality, the angular momentum of the cloud becomes
nonnegligible  before then and the mass of the fragments is
determined by criteria other than eq.\ (\ref{eq:jeansmass}).

We focus on 
the cloud collapse inside the  MW halo at $z=10$, which
according to the merger trees has an  average mass  $M\simeq 8
\times10^8$ M$_\odot$ and central density  $n\sim1$ cm$^{-3}$.  It is
assumed that the collapse can always  happen, independently of $x$ and
$\beta$, and that the fragments can cool to  $\sim10$~\% of $T_M$ (see
eq.\ (\ref{eq:virTemp})) before collapsing. In section \ref{sect:mgf}
we will verify the values of $(x,\beta)$ for which cooling is really
efficient enough for the cloud to collapse. In such a case $T$
drops  to values $\ll T_M$ before the density increases significantly.
Then our assumptions are self-consistent and allow for estimating the
mass of primordial stars independently of $\beta$; dependence on $x$
remains since it affects the chemical abundances.

\begin{figure*}[t]
	\begin{center}
		\includegraphics[width=0.5\linewidth]{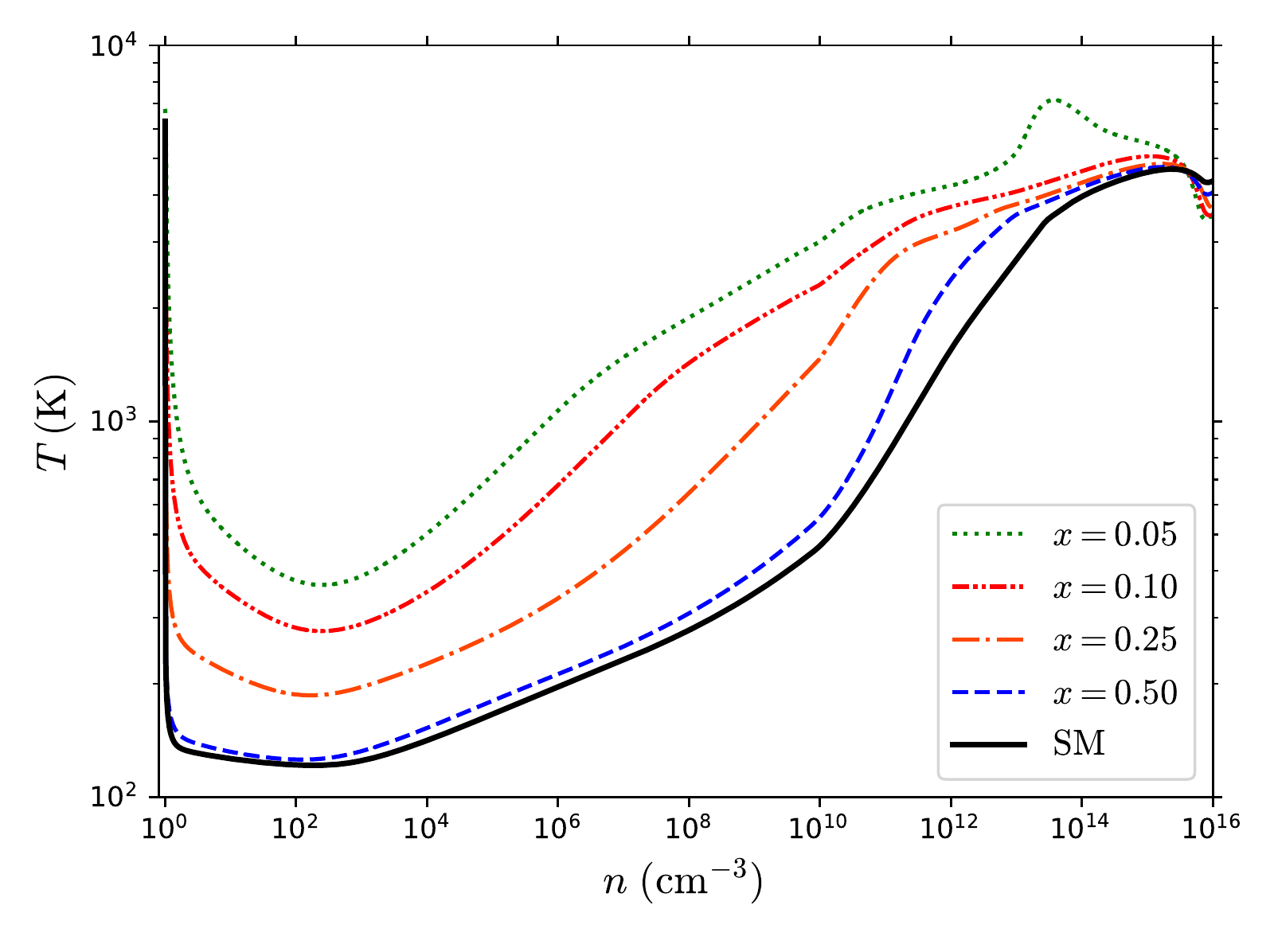}\hfil
		\includegraphics[width=0.5\linewidth]{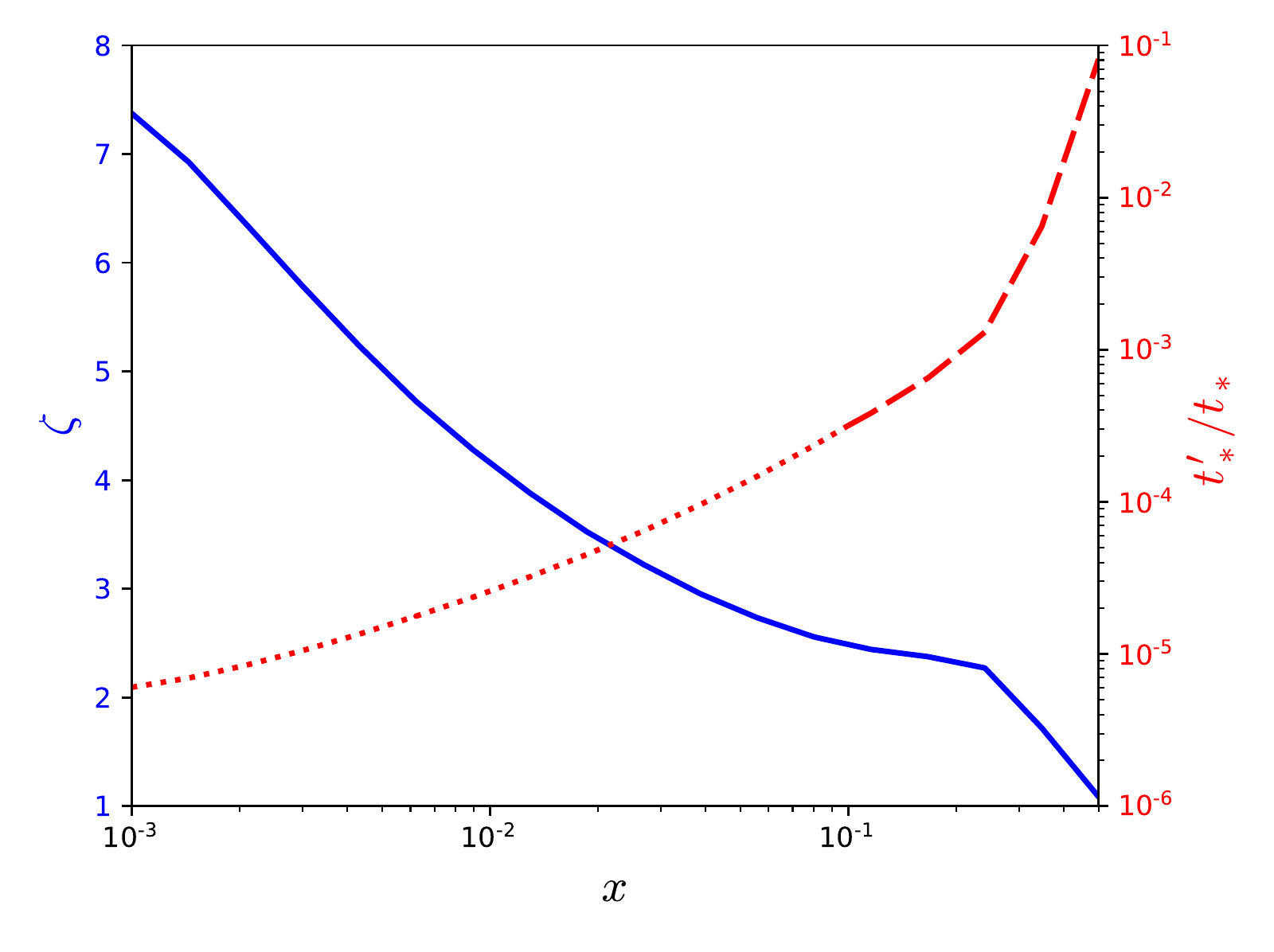}
		\caption{Left: Temperature evolution during the cloud collapse of a 
			gas fragment at $z=10$ in a Milky Way-like
			halo for 
			values of $x$ increasing from top to bottom, 
	including the SM ($x=1,Y=0.24$). Evaluating eq.\ 
	(\ref{eq:jeansmass}) 
			at 
			the temperature minimum of each curve gives an
estimate of the mass of 
			primordial 
			stars. Right: Ratio $\zeta$ of the minimal fragment mass in the 
	\MS\
			relative to the SM (blue, solid) and ratio of the 
			characteristic stellar 
			lifetimes (red, dashed+dotted). The dotted curve 
			illustrates 
			the extrapolation of eq.\ (\ref{eq:tstar}) outside the fit interval 
			of ref.\ \cite{Berezhiani:2005vv}.}
		\label{fig:mirrorStars}
	\end{center} 
\end{figure*}

Fig. \ref{fig:mirrorStars} shows the evolution of the temperature from
eq.\ (\ref{Tevol}) during  the  collapse for several values of $x$ and
for the SM. It reveals that smaller values of $x$ lead to more
efficient cooling, since more H$_2$  can form. Interestingly, even
when the hydrogen fraction is small, H$_2$  cooling can reduce $T$ to
$\sim {\rm few}\times 100$\,K very rapidly. We  evaluated the
Jeans mass at the minimum  $T$ (near  $n\sim 200$
cm$^{-3}$) to estimate the mass $M_*'$ ($M_*$) of the fragments in 
the \MS\ (SM). After this point, the cloud collapses quasi-adiabatically and 
the rise of $T$ slows the decrease of the Jeans mass. This point allows us to 
set an upper limit on the final fragment mass $M_*'$ rather than evaluating it 
accurately, which is 
impossible in our simplified analysis without angular momentum.

 Note that this oversimplification 
	is not an issue, because we are only interested in the ratio $\zeta\equiv 
	M_*'/M_*$ of the fragment mass  in the \MS\ and in
	the SM,	which wouldn't change much if we evaluated eq.\ 
	\ref{eq:jeansmass} at another point of the $n\!-\!T$ diagram. This ratio  
	gives a rough 	approximation of how the mass of the
	mirror stars scales compared  to the visible ones, which allows us to
	estimate their lifetimes and their  supernova feedback on structure
	formation.
Fig.\ \ref{fig:mirrorStars} also illustrates the value of $\zeta$ for all 
values of $x$.  It is apparent that $\zeta>1$ for all  values  of $x$, 
indicative of the lower cooling efficiency in the
\MS\ (from suppressed H abundance) leading to less  fragmentation
of gas clouds.

We should emphasize that this estimate of $\zeta$ is only valid for 
primordial stars as we do not include any element heavier than He in our 
analysis. In reality, it is possible that the short-lived He-dominated stars in 
the \MS\ produce metals at a much higher rate than in the SM. Since metals are 
easier to ionize, their presence can significantly increase the cooling rate 
and the fragmentation inside a gas cloud. We leave this analysis for a 
future study.

Unlike ordinary matter, mirror stars are usually He-dominated,
which  has important consequences for their evolution, notably
their  lifetime $t_*$. In the SM, the
H-burning phase constitutes most of the lifetime of stars, with
the post-main sequence evolution  contributing only about 10~\% of
$t_*$. In the \MS, with much  less H to burn, stars
quickly transition to the later stages of their  evolution. 

We note that the average mass for 
visible stars can be estimated using the initial mass function (IMF):
\be
M_* = \int_{0.08\,M_\odot}^{100\,M_\odot} m\, \phi(m)\, dm \simeq 0.3\ M_\odot,
\ee
where $\phi(m)\propto m^{-2.35}$ is the Salpeter IMF 
\cite{MBW:2010,1955ApJ...121..161S} 
normalized such that its integral over the mass range of stable stars ($0.08\ 
M_\odot<m<100\ M_\odot$) is 1. Hence we take the characteristic  stellar mass 
in the \MS\ to be $M_*' = \zeta\times(0.3$ M$_\odot)$. Ref.\ 
\cite{Berezhiani:2005vv} studied the dependence of $t_*$ on the
He fraction and the mass of stars.  Using their fit results, we estimate the 
scaling of typical lifetimes of \MS\ stars  by comparison to the SM: 
\begin{align} \label{eq:tstar}
\log_{10}\left(\frac{t_*'}{t_*}\right) & \simeq 0.74 - 2.86 Y' -0.94
Y'^2 -  4.77 \log_{10} \zeta\nn\\ & +0.99 (\log_{10}\zeta)^2 + 1.34 Y'
\log_{10}\zeta  \\ &+ 0.29Y'^2\log_{10}\zeta -0.28 Y'^2 
(\log_{10}\zeta)^2.\nn \end{align}
(recall that $\zeta=M'_*/M_*$). 

The ratios $\zeta$ and $t'_*/t_*$ are plotted in figure
\ref{fig:mirrorStars}(right) as a function of $x$.
$t_*'/t_*$ will be used to estimate the  supernova
feedback of \MS\ stars on the formation of dark galactic structures 
in section \ref{sec:disk}.  We note that eq.\ (\ref{eq:tstar}) is only valid 
up to $Y'= 0.8$ ($x\simeq 0.1$), so our estimate of the stellar lifetime 
for $x\lesssim0.1$ is likely to be too small. However, in this range one
nevertheless expects that $t'_*/t_*\ll1$. In section \ref{sec:disk} we will show
that the main 
consequence of such short lifetimes is that supernova feedback 
favors star production over the formation of cold gas clouds in the mirror 
galactic disk, whereas in section \ref{sec:constraints} we show that star 
formation is already maximally efficient at $x=0.1$; hence our results our
not sensitive to 
the precise value of 
$t'_*/t_*$ at lower temperatures, and it is safe to 
use eq.\ (\ref{eq:tstar}) for $Y'>0.8$. 

\subsection{Mirror galaxy formation}
\label{sect:mgf}

We now have the necessary ingredients for studying  the formation of
a dark galaxy using the GALFORM model \cite{Cole:2000ex}. The steps
to be carried out for implementing it are described as follows.

The \MS\ matter is divided into three components: the hot gas
component,  the spheroidal bulge fraction and the disk fraction. The
bulge and  the disk together form the \MS\ galaxy. The disk  fraction
is further subdivided into two components: active stars and cold gas
clouds. Star formation  is highly suppressed in the bulge so such a
subdivision is not needed there.
The remaining matter component is CDM. Visible baryons are omitted
from our analysis for simplicity and since GALFORM is not set up to
properly account for their gravitational interaction with the 
\MS.\footnote{The only impact of SM particles in our analysis would be
	to  potentially shorten the free-fall timescale, eq.\ (\ref{eq:tff}) by
	collapsing  and changing the total matter distribution in the halo.
	First, since visible  baryons only represent about 15~\% of the total
	matter content, their impact on  $t_{\mathrm{ff}}$ is small.
	Secondly, structure formation in the \MS\ also  equally depends on the
	cooling timescale, eq.\ (\ref{eq:tcool}), which is  independent of the
	SM matter.}\ \   Instead, we include the visible baryons into the CDM 
	fraction, so that $\Omega_m = \Omega_c + \Omega_{b'}$.

CDM is assumed to have an NFW density profile,

\be \label{eq:NFW} 
\rho\cdm (r) = \frac{ (\Omega_c/\Omega_m) 
(\Delta\vir\rho_{\mathrm{crit}}/3)}{
\left[\ln(1+c)-\cfrac{c}{1+c}\right]  \cfrac{r}{r\vir} \left(\cfrac{r}{r\vir} 
+\cfrac{1}{c}\right)^2}.
\ee 
The virial radius $r\vir$ and overdensity $\Delta\vir$ were defined
in  section \ref{virial}. $c$ is the NFW concentration and  sets the
size of the central region of the profile. The  procedure to find $c$ 
for a given halo mass at a given redshift is described in the appendix
of ref.\ \cite{Navarro:1996gj}. The CDM profile remains constant 
throughout the lifetime of the halo.

\MS\ matter is further assumed to form a hot gas cloud with an 
isothermal density profile,
\be\label{eq:mirrorProfile}
\rho_{b'} (r) = \frac{f_{\mathrm{hot}} (\Omega_{b'}/\Omega_m) 
	(\Delta\vir\rho_{\mathrm{crit}}/3)}{  \left[1 
	- 
	\cfrac{r_0}{r\vir} \tan^{-1}\left(\cfrac{r\vir}{r_0}\right)\right]
	\left(\left(\cfrac{r}{r\vir}\right)^2+\left(\cfrac{r_0}{r\vir}\right)^2\right)}.
\ee 
The hot gas fraction $f_{\mathrm{hot}}=1$ for all newly formed halos, but 
$f_{\mathrm{hot}}$ decreases as the gas cools and collapses. The core radius 
$r_0$ is 
initially related to the NFW concentration as $r_0/r\vir = 1/(3c)$, 
but as the gas cools, $r_0$ increases such that the 
density and pressure at $r\vir$ remain unaffected. This is impossible in the 
limit where a large fraction of the gas cools, so we set an upper limit of 
$r_0=15 r\vir$ to avoid a numerical divergence as $r_0\to \infty$. In this 
limit $(r/r\vir)^2\ll (r_0/r\vir)^2$ and the density profile becomes 
essentially homogeneous. We truncate both profiles $\rho_{b'}(r)$ and 
$\rho_c(r)$ at $r\vir$.

\subsubsection{Disk formation}
\label{sec:disk}

The GALFORM algorithm simulates structure formation beginning at 
redshift $z=10$, taking as input the merger trees described above, 
and evolving forward in time using logarithmically spaced time steps
$\Delta t$.  Halo evolution is simulated semi-analytically until the
present, $z=0$.  The lifetime of the halo $t_{\mathrm{halo}}$ is
defined to be the time it takes to double in mass, whether by matter
accretion or by mergers.

The halo is modeled using spherical shells plus a disk
component.  At the beginning (or end) of each time step, two characteristic 
radii must be
computed: the cooling radius
$r_{\mathrm{cool}}$ and  the free-fall radius $r_{\mathrm{ff}}$.
These are respectively the
maximal distances such that the cooling timescale
$t_{\mathrm{cool}}$ (eq.\  (\ref{eq:tcool})) and the free-fall
timescale (eq.\ (\ref{eq:tff})) are smaller  than the elapsed time since
the beginning of the  halo's lifetime, $t-t_{\mathrm{halo}}$.
Hence the radius  $r_{\mathrm{acc}} = \min(r_{\mathrm{cool}},
r_{\mathrm{ff}})$ is the maximum  distance to which the gas has had
time to cool down and accrete into compact objects.

The values of $r_{\mathrm{acc}}$ before and after the time step
$\Delta t$  delimit a  spherical shell of width $\Delta
r_{\mathrm{acc}}$ that contains mass  $\Delta M_{\mathrm{acc}}$ of
hot mirror matter gas. As shown in appendix B of  ref.\
\cite{Cole:2000ex}, this accreted matter determines how the masses of
the  hot gas $M_{\mathrm{hot}}$ and the disk $M_{\mathrm{disk}}$
change during that  time step:

\begin{align}
	\Delta M_{\mathrm{disk}} &= \Delta M_{\mathrm{cold}} + \Delta M_* 
	\label{eq:disk1}\\
	\Delta M_{\star}&=M_{\mathrm{cold}}^{0} 
	\frac{1-R}{1-R+B}\left[1- e^{-\Delta t / 
	\tau_{\mathrm{eff}}}\right] \\	
	\nn&-\Delta M_{\mathrm{acc}} \frac{\tau_{\mathrm{eff}}}{\Delta t}  
	\frac{1-R}{1-R+B}\left[1-\frac{\Delta t }{\tau_{\mathrm{eff}}}	- 
	e^{-\Delta 	t / \tau_{\mathrm{eff}}}\right]	\\
	\Delta M_{\mathrm{cold}}&=\Delta M_{\mathrm{acc}}-\frac{1-R+B}{1-R} 
	\Delta M_{\star} \\
	\Delta M_{\mathrm{hot}}&=-\Delta M_{\mathrm{acc}}+\frac{B}{1-R} \Delta 
	M_{\star}.\label{eq:disk4}
\end{align}
 Here $ M_{\mathrm{cold}}$ and $M_*$ are the masses of the cold 
gas and stellar components of the disk, and $M_{\mathrm{cold}}^{0}$ is the cold 
gas 
mass at the beginning of the time step. $R$ is the fraction of mass recycled by 
stars ({\it e.g.,} stellar winds that contribute to the cold gas component of 
the 
disk) and $B$ parametrizes the efficiency of the supernova feedback that heats 
the cold gas fraction. 

The effective mirror star formation timescale
is  $\tau_{\mathrm{eff}} = \tau_*' / (1-R+B)$.  To determine
$\tau_*'$, one can assume that the star formation rate is in equilibrium with 
the stellar death rate 
(the inverse of the average stellar lifetime).  Then the ratio of star 
formation 
timescales $\tau_* '/\tau_*$ in the \MS\ and in 
the SM is equal to the ratio of the characteristic stellar lifetimes 
$t_*'/t_*$, eq.\ (\ref{eq:tstar}),
\be  \label{eq:tauStar}
\tau_*' \simeq \left(\frac{t_*'}{t_*}\right) \tau_* = 200  \frac{r_D}{V_D} 
\left(\frac{t_*'}{t_*}\right) \left(\frac{V_D}{200\ \mathrm{km/s} 
}\right)^{-1.5}.
\ee

Following ref.\ \cite{Cole:2000ex}, we take  $R=0.31$ and $B
=\left(V_{\mathrm{D}}/( 200\  \text{km/s})\right)^{-2}$, where
$V_{\mathrm{D}} = (G M_{r_D}/r_D)^{1/2}$ is  the  circular velocity at
the half-mass radius $r_D$ of the galactic disk. Assuming  the disk
has an exponential surface density, its half-mass 
radius can be estimated as $r_D = 1.19 \lambda_H r_{\mathrm{acc}}$ where 
$\lambda_H$ is a spin  parameter that follows a log-normal distribution with
average value $\lambda_H=0.039$, that we adopt for simplicity.

The evolution of the disk and the hot gas mass fractions is found by
iterating eqs.\ (\ref{eq:disk1}-\ref{eq:disk4}).  During the
characteristic time $t_{\rm halo}$, the temperature $T_M$ of the hot 
mirror matter gas is assumed to  remain at its initial value, eq.\
(\ref{eq:virTemp}), and likewise for the relative abundances of each
chemical species and the core radius $r_0$  of the hot gas density
profile.  All of these quantities are updated at the beginning of each
stage of evolution spanning time $t_{\rm halo}$, for all the active
halos of the  merger tree.

\subsubsection{Galaxy mergers}

Eventually, every halo in the merger tree combines with another halo,
the smaller of the two becoming a satellite of the larger
one. We assume that  all the hot gas of the satellite halo is stripped
by hydrodynamic drag,  so that its disk and bulge fractions no longer
evolve. After this the satellite orbits the main halo
until they merge, over the characteristic timescale
\be 
\tau_{\mathrm{mrg}} = \Theta_{\mathrm{orbit}} \frac{\pi r\vir}{V_H} 
\frac{0.3722}{\ln (M_H/M_{\mathrm{sat}})} \frac{M_H}{M_{\mathrm{sat}}}.
\ee 
Here $V_H= (GM_H/r\vir)^{1/2}$ is the circular velocity at the virial radius, 
$M_{\mathrm{sat}}$ is the total mass of the satellite halo (mirror baryons and 
CDM) and $M_H$ is the total mass of the main halo, including all the satellite 
halos. $\Theta_{\mathrm{orbit}}$ is a parameter that depends on the orbit of 
the satellite. It is characterized by a random log-normal distribution 
with an average $\langle\Theta_{\mathrm{orbit}}\rangle =e^{-0.14}$ and a 
standard deviation $\sigma =0.26$.

The outcome of a galaxy merger depends on the mass ratio of the two
galaxies  (disk and bulge components only), $M_{\mathrm{gal}}^{\
	\mathrm{sat}} /  M_{\mathrm{gal}}^{\ \mathrm{cen}} $. If this ratio is
smaller than a critical  value $f_{\mathrm{crit}}$, the merger is
``minor:'' the satellite galaxy is  disrupted, its bulge and stellar
components are added to the bulge  fraction of the central galaxy,
and the cold gas falls into the central  disk. If the mass ratio is
greater than $f_{\mathrm{crit}}$, the merger is  ``major,'' in which
case both galaxies are disrupted by dynamical friction and  all the
mirror matter ends up in a spheroidal bulge. We take
$f_{\mathrm{crit}}=0.3$, the lowest possible value in
agreement with  numerical studies \cite{Cole:2000ex}, but it has
been argued in ref.\  \cite{Ghalsasi:2017jna} that larger values
do not change the results  significantly.

In a minor merger, the cold gas of the satellite galaxy is added to
the main  galactic disk, which changes its half-mass radius $r_D$.
The new  radius is determined by the conservation of angular momentum
$j_D = 2 r_D V_H/1.68$.  Averaging over the relative orientation of
the two galaxies yields
\be 
r_{Df} = \frac{r_{D1}M_{D1} + r_{D2}M_{D2}}{M_{D1}+ M_{D2}},
\ee
that is, the new radius is the weighted average of the two initial
radii. The  bulge component is expected to have a de Vaucouleurs density 
profile,
$\log  \rho_{\mathrm{bulge}}\sim - r^{1/4}$, but we find that it can 
be more simply modeled as a sphere of uniform density and radius
$r_D/2$, without significantly changing the final results.

By iterating over all halos and evolving until 
$z=0$, the  procedure described in this section allows us to predict
the fraction of mirror  matter that forms galactic structures (either
a disk or a bulge) and the  fraction that remains in a hot gas cloud.
We simulated galaxy evolution in 10 different merger trees for $18^2$ 
combinations of $(x,\beta)$ in the range 
$10^{-3} < x < 0.5$ and $10^{-3} < \beta < 5$ and averaged over the final 
fractions. Smaller values of $\beta$ cannot be constrained with
present data given the current experimental sensitivity to a
very subdominant component of \MS\ dark matter.
Similarly, for $x<10^{-3}$ the helium mass fraction 
is saturated ($Y'\sim 0.99$) and the chemical evolution of the \MS\
gas cannot be distinguished from that at $x=10^{-3}$.
In section \ref{sec:constraints}, we will use these predictions in conjunction 
with
astronomical data, to  constrain the parameters of the
model. 

\section{Constraints on mirror dark matter structures}
\label{sec:constraints}

\begin{figure*}[t]
	\begin{center}
		\includegraphics[width=0.5\linewidth]{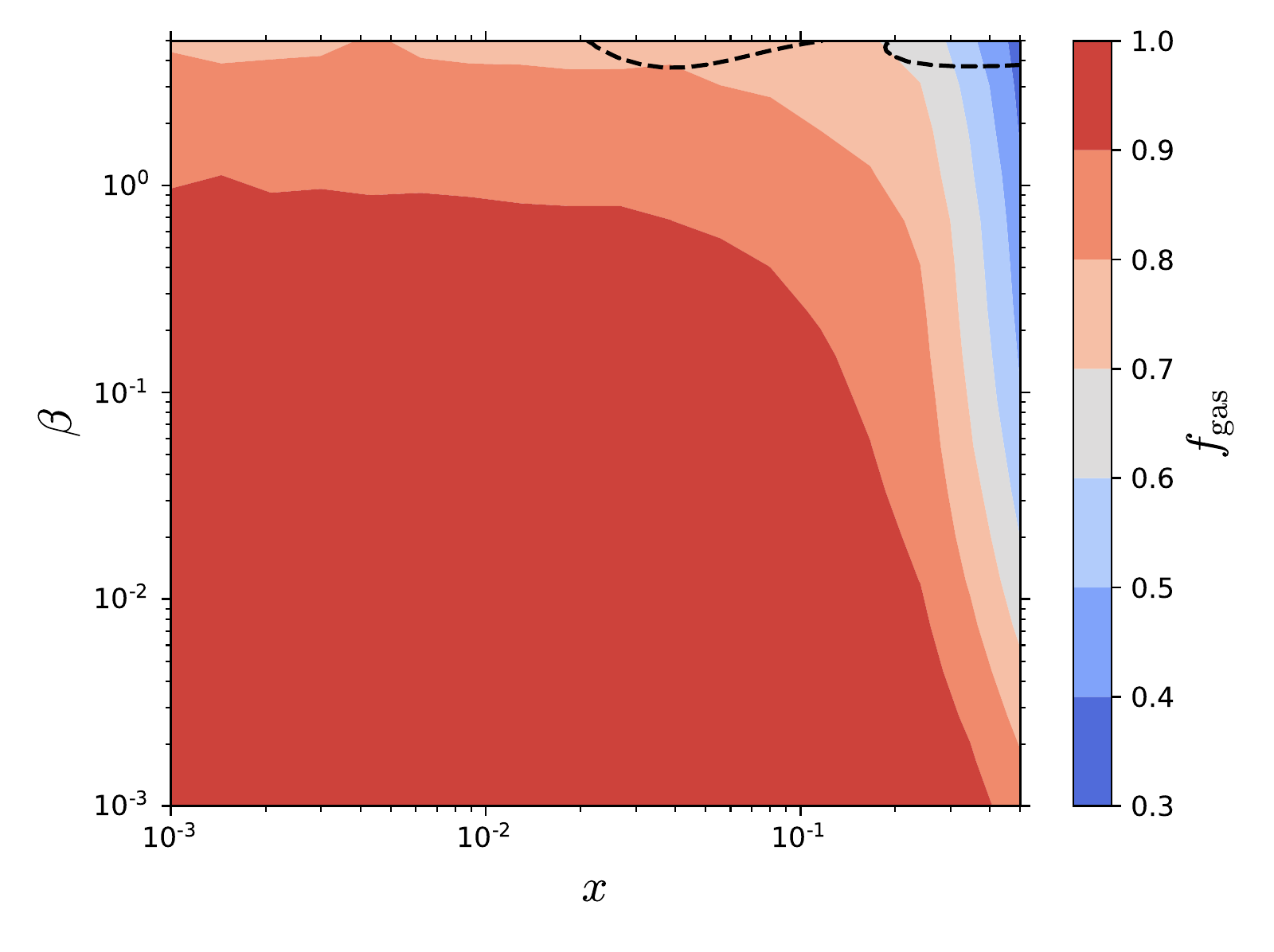}\hfil
		\includegraphics[width=0.5\linewidth]{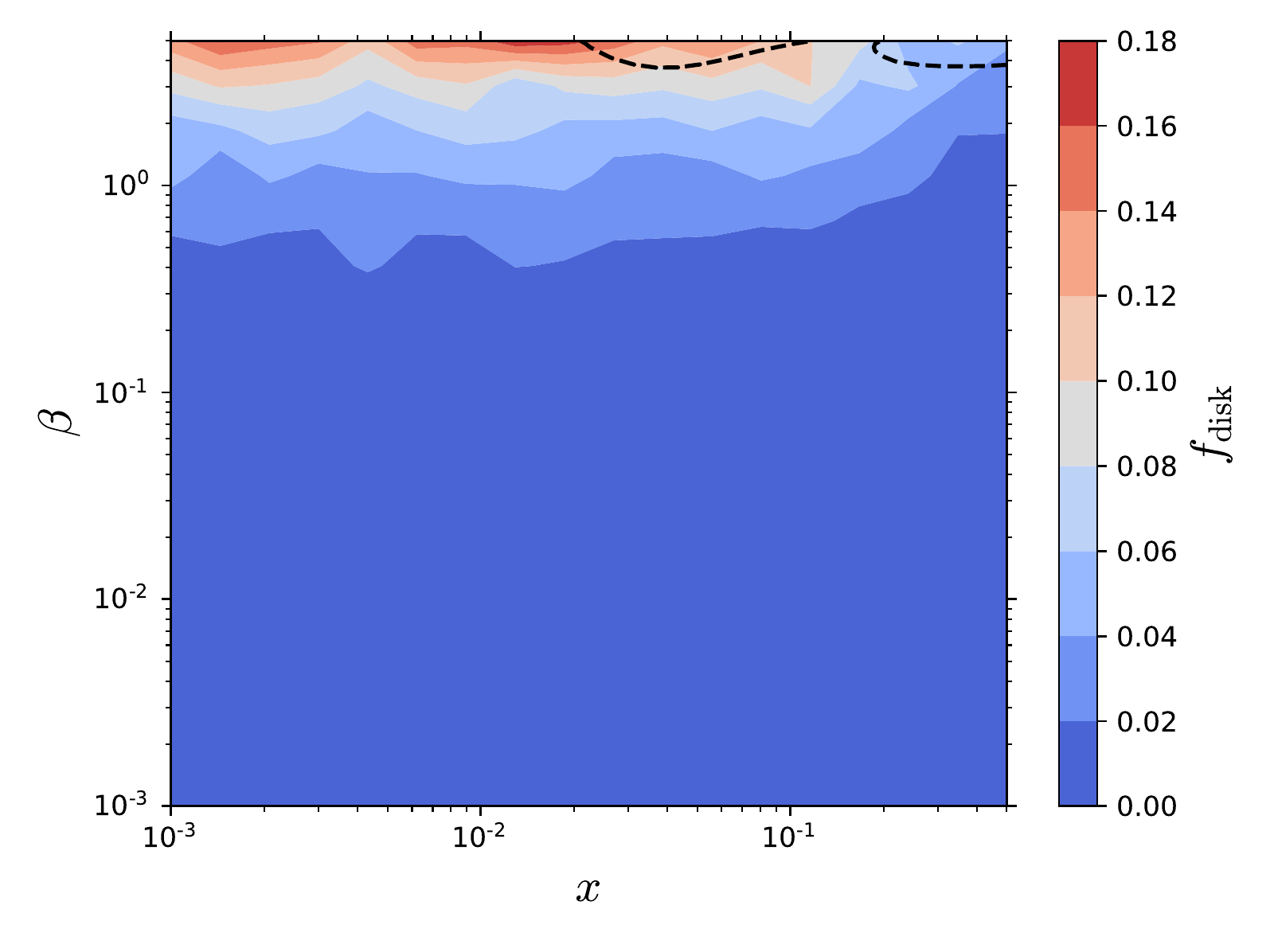}\\
			\includegraphics[width=0.5\linewidth]{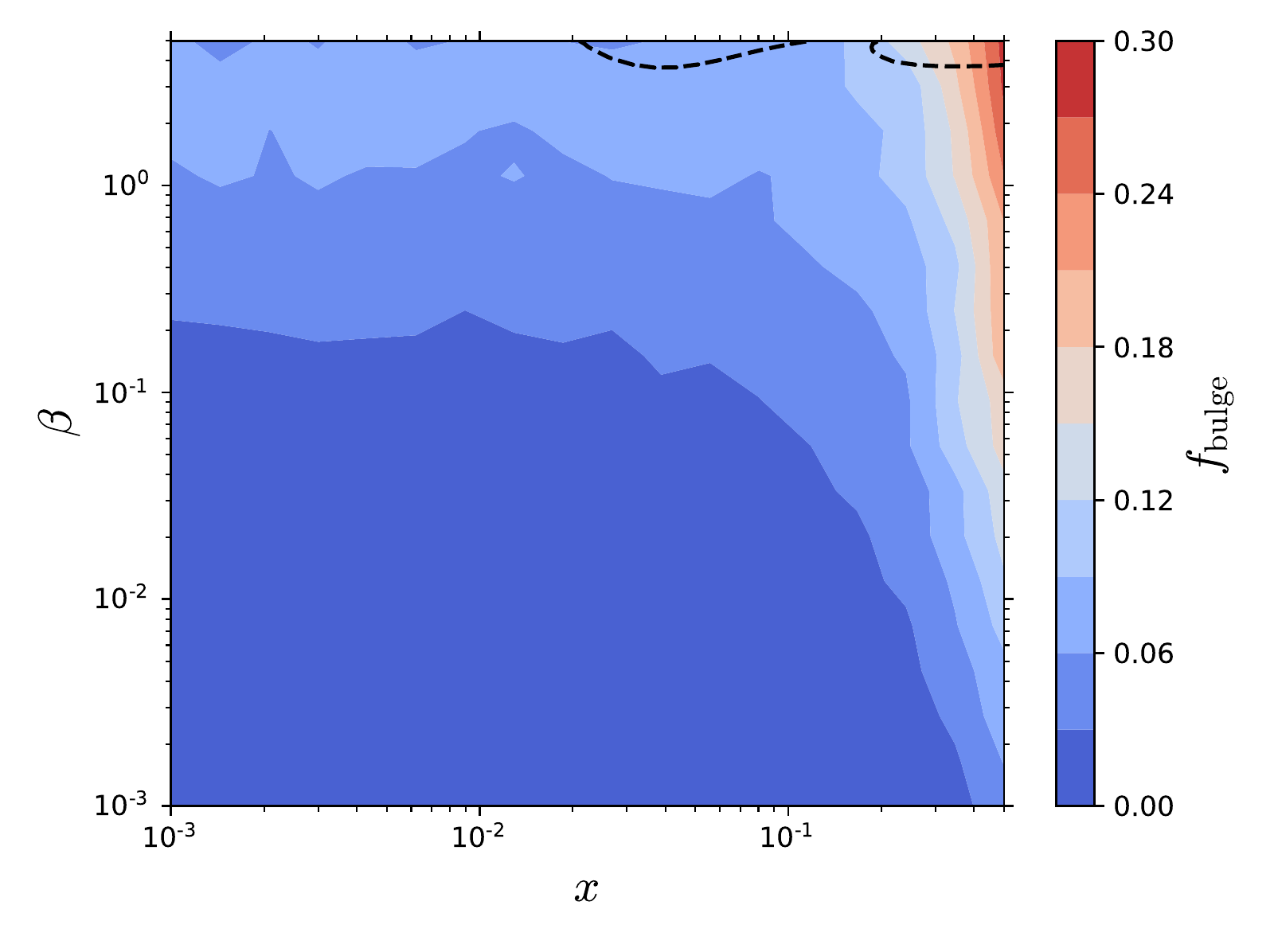}\hfil
		\includegraphics[width=0.5\linewidth]{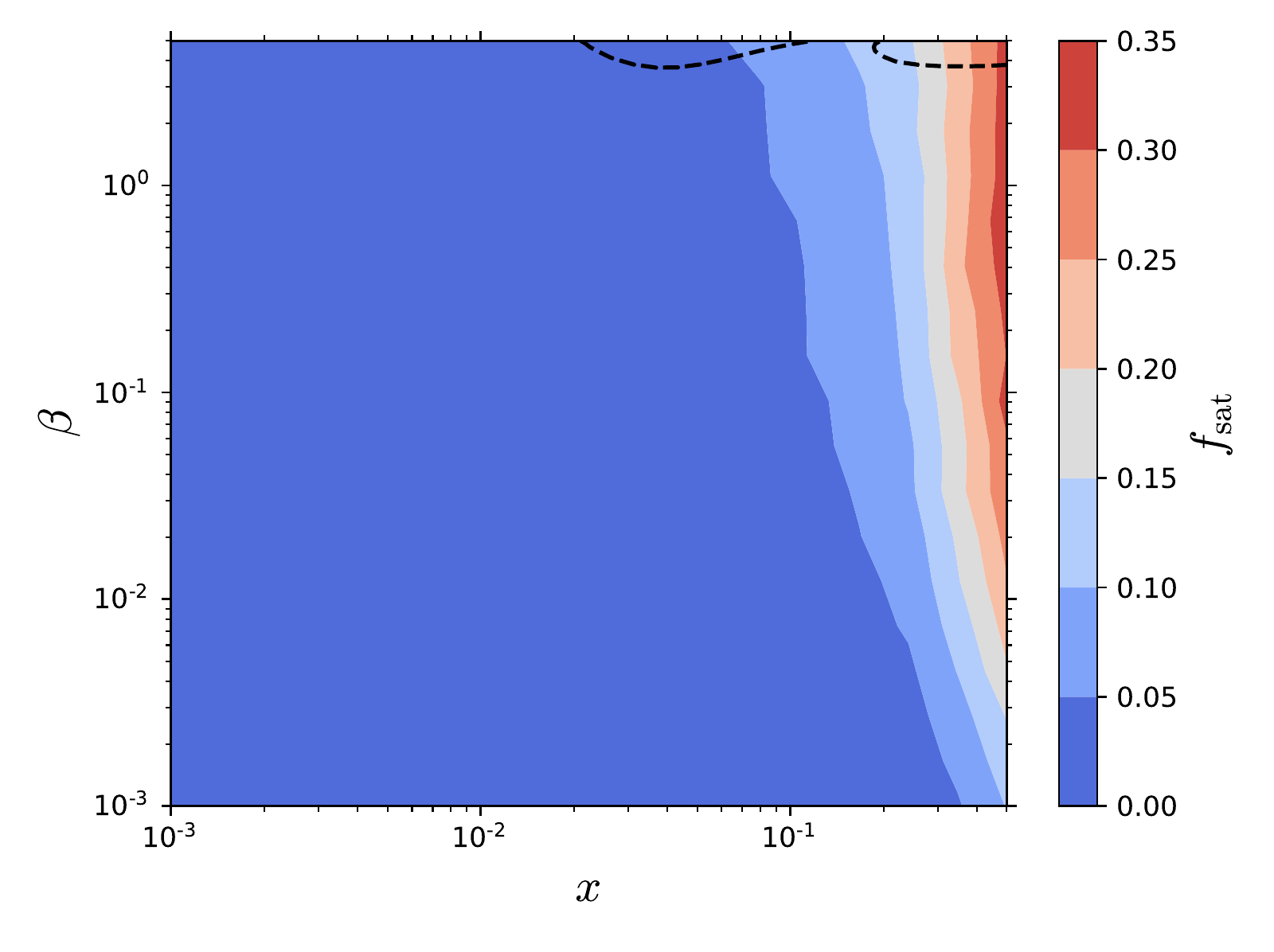}\\
			\includegraphics[width=0.5\linewidth]{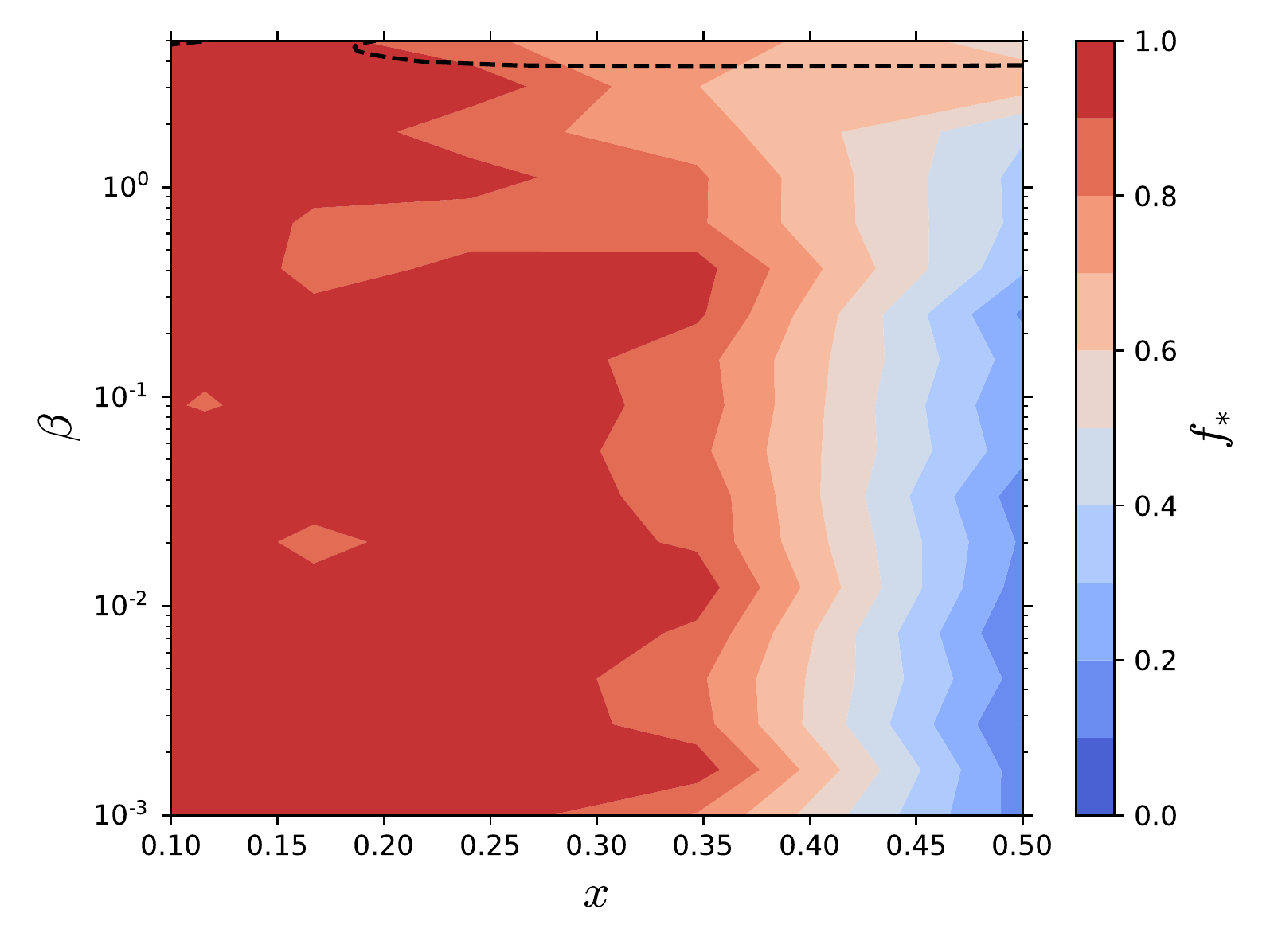}
					\caption{Results of the mirror structure formation 
					analysis. The top four panels show the average fraction of 
					mirror particles in each
component of galactic structure (hot gas, 
					disk, bulge and satellite galaxies) in a 
					$10^{12}$~M$_\odot$ halo such that 
					$f_{\mathrm{gas}}+f_{\mathrm{disk}} + 
					f_{\mathrm{bulge}} + f_{\mathrm{sat}}=1$. The bottom panel 
					shows the fraction of stars
$f_*$ in the mirror galactic 
					disk. The fraction of cold gas in the disk is given by 
					$f_{\mathrm{cold}} = 1-f_*$. 
					The regions above the dashed curves 
					are excluded from our analysis due to the self-consistency 
					check discussed in section \ref{sec:Silk}.}
		\label{fig:results}
	\end{center} 
\end{figure*}

The results of our \MS\ structure formation analysis are shown in figure 
\ref{fig:results}, where the fractions of the different components
$f_{\mathrm{gas}}$, $f_{\mathrm{disk}}$, $f_{\mathrm{bulge}}$, $f_{\mathrm{sat}}$
and $f_*$ (the fraction in stars in the disk) are plotted as functions of $(x,\beta)$.
One of the most striking features is that for much of the parameter space 
($x\lesssim0.1$, $\beta\lesssim 1$), over 90~\% 
of mirror matter is in a hot gas cloud and does not condense to form structures 
in the halo. This is readily understood, since the low density and low hydrogen 
abundance lead to inefficient cooling, maintaining high pressure 
in the gas cloud and preventing it from collapsing. Our results show that 
at low $x$ and in the range $0.5\lesssim \beta \lesssim 1$ about 5--10~\% of 
the \MS\ forms a dark galaxy. In this case, even if the dark galaxy is 
subdominant in the halo, the mirror stars and supernovae within it would 
amplify the baryonic effects of SM particles,  which have been argued to significantly 
alleviate the small-scale tension of CDM \cite{2014ApJ...786...87B}.

For \MS\ densities $\beta\lesssim0.5$, mirror matter behaves similarly to generic models 
of dissipative DM, such as atomic DM, that have no  nuclear or chemical 
reactions and do not collapse into compact objects. Although the \MS\ 
would constitute only a small fraction of DM and would not lead to dark 
stuctures (stars, planets, life forms), its dissipative effects could still have 
interesting cosmological effects, like the suppression of the matter power spectrum on small 
scales. The mirror gas cloud would also have a cored density profile, resulting 
in a shallower gravitational potential in the center of the halo than in a pure 
CDM scenario, possibly ameliorating the cusp-core problem.

The disk fraction $f_{\mathrm{disk}}$ depends much more strongly on 
$\beta$ than on $x$. This comes about because the long 
lifetime of the main halo allows for the formation of a mirror 
galaxy at sufficiently high density, even though cooling is less 
efficient at small $x$ (due to the low hydrogen fraction). 
The fraction 
$f_{\mathrm{sat}}$ of mirror matter in satellite galaxies behaves differently: 
even at large mirror particles densities, for $x<0.1$ the cooling timescale becomes 
longer than the lifetime of subhalos merging with the MW, leaving too little time
for structures to form. Hence 
dwarf galaxies orbiting the MW will host few mirror particles 
if $x<0.1$.   It is likely however that we underestimate
$f_{\mathrm{sat}}$ due to our assumption that galaxy formation 
ended once the subhalos 
merged with the main halo. In reality the satellite galaxies can accrete 
cooling gas from the main halo and continue to grow after a merger. 

There is a clear correlation between $f_{\mathrm{bulge}}$ and the sum of 
$f_{\mathrm{disk}}$ and $f_{\mathrm{sat}}$, which arises because
bulge formation requires both the main halo and the satellite subhalos to 
form, before the latter is disrupted by dynamical friction. The 
absence of a disk for $x\gtrsim0.1$, where $f_{\mathrm{bulge}}$ is at its 
maximum, indicates that a major merger destroyed the disk of the central 
halo. That major merger is probably recent, otherwise the disk would have 
had time to form again. Similarly, we can understand the small bulge 
fraction in the region $\beta\gtrsim0.2$, $x\lesssim0.1$ as resulting from a 
series of minor mergers or an early-time major merger,
since there is a significant disk fraction at $z=0$ for these parameters.

The bottom panel of fig.\ \ref{fig:results} shows the effect of the shortened 
stellar lifetime in the \MS\ (see eqs.\ (\ref{eq:tstar}) and 
(\ref{eq:tauStar})). 
The high He abundance and the larger mass of primordial stars increase the 
stellar feedback from supernovae to a point where most of the cold molecular 
clouds are rapidly heated and return to the hot fraction of the halo, 
leaving mirror stars as the only inhabitants of the mirror galaxy.

\begin{figure}[t]
	\begin{center}
		\includegraphics[width=\linewidth]{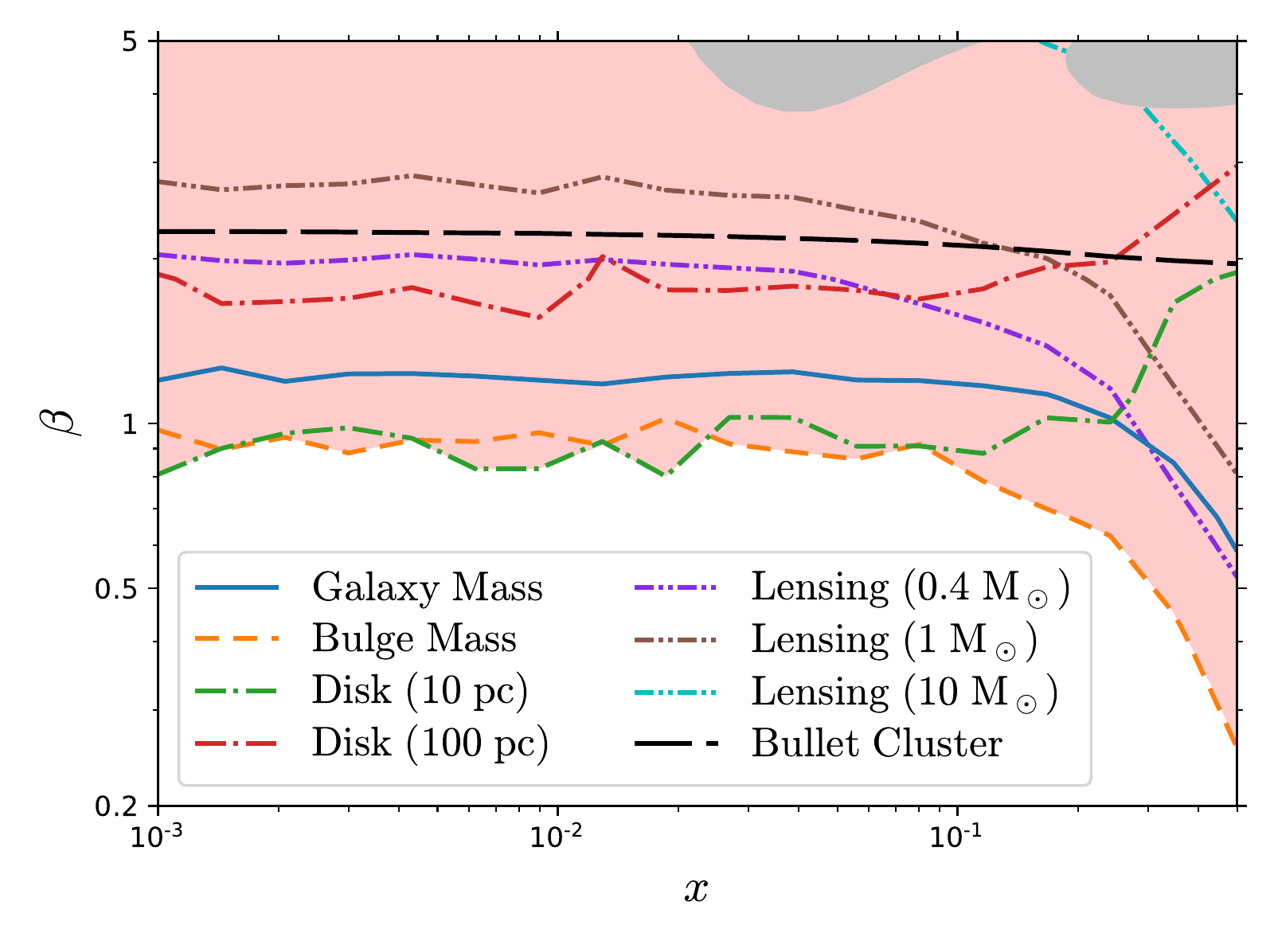}
		\cprotect\caption{Upper limits for \MS\ model from 
		constraints on: 
		the total mass of the galaxy (defined as bulge plus disk, solid curve); 
		the bulge mass (dashed); 
	the thin disk surface 
		density (dot-dashed), assuming $h_D = 10$ (green) or 100 pc (red) 
for the dark disk; gravitational lensing 
		events (double-dot-dashed) for $m_{\mathrm{mac}}=0.4$ (violet), 
		1 (brown) or 10 (cyan) M$_\odot$; and the Bullet Cluster (long-dashed). 
		The red shaded area is excluded, while 
 the grey regions lie outside the validity of
		our analysis (see section \ref{sec:Silk}).}
		\label{fig:constraints}
	\end{center} 
\end{figure}

Next we consider various astronomical constraints on \MS\ galactic 
structures. The excluded regions lie above the curves shown in fig.\ 
\ref{fig:constraints}.   The limits on disk surface density, bulge and total stellar
mass, and from gravitational lensing surveys, are described in the following.

\subsection{Thin disk surface density}

Data from Gaia DR2 allowed ref.\ \cite{Buch:2018qdr} to constrain the 
surface density $\Sigma_D$ of a thin dark disk in the vicinity of the Sun. 
A gravitational potential in the presence of a DM disk would 
be deeper, leading to greater acceleration towards the galactic plane than 
what ordinary stars can account for. This affects the transverse velocities 
and 
density distribution of nearby stars. Assuming that the dark disk possesses an 
exponential profile and a scale height $h_D\simeq 10$~pc (which could explain 
phenomena like the periodicity of comet impacts 
\cite{r2014dark,schutz2017constraining}), the 95~\% C.L. bound 
on its local surface density is 
 \be \label{eq:diskBound}
 \Sigma_D (R_\odot) = \frac{M_D}{2 \pi L_D^2} e^{-R_\odot/L_D} \lesssim 4.15\ 
 {\text{M}_\odot\over\text{pc}^2},
 \ee 
where $R_\odot=8.1$ kpc is the distance of the Sun from the center of the 
galaxy and $L_D$ is the scale length of the disk. The scale length is related 
to the half-mass radius $r_D$, which we included in our analysis, as
$r_D/L_D \simeq 1.68$.

The constraint (\ref{eq:diskBound}) led ref.\ \cite{Buch:2018qdr} to conclude that a
dissipative dark sector can constitute less than 1~\% of the  total DM. However 
a more
conservative interpretation is  that less than 1~\% of the DM \textit{has 
accreted into a
thin dark disk}; in that case the  dissipative dark sector could be more abundant
since we expect only a  fraction of it to form a galactic disk, $\lesssim20~\%$ 
for
mirror matter, as shown in fig.\  \ref{fig:results}. 

Assuming a thin disk with scale height $h_D=10$\,pc, this bound rules out the region 
$\beta\gtrsim 1$, except for $x\gtrsim0.25$ where it is relaxed to
$\beta \gtrsim1.8$. For a thicker disk with $h_D=100$\,pc,
closer to the height of the visible disk, the 
constraint is relaxed to $\Sigma_D (R_\odot)  \lesssim 12.9\ 
\text{M}_\odot/\text{pc}^2$, loosening the bound on $\beta$ by a 
factor of $\sim2$.

An underlying assumption is that the 
dark disk lies withing the MW plane. Although the two disks need not be 
initially aligned, one expects their gravitational 
attraction to do so on  
the dynamical timescale of the inner region of the halo, $t_{\mathrm{dyn}}\sim 
1/\sqrt{G\rho} \sim 
\sqrt{L_D^3/G M_D}$. Even if the dark disk has
a negligible density such that only the visible disk contributes to  
$t_{\mathrm{dyn}}$ 
($M_D\sim10^{10}$~M$_\odot$, $L_D\sim 2.5$~kpc), 
one finds $t_{\mathrm{dyn}}\sim 20$\,Myr, much shorter than the lifetime of the halo. Hence in all 
cases the two disks should be coincident.

\subsection{Bulge and total stellar mass}

Data from Gaia DR2 further enabled ref.\ \cite{cautun2019milky} to determine the total mass of each 
component 
of the MW halo by fitting the rotation curves of nearby stars and using other 
kinematical data. They determined the mass of the galaxy (disk 
and bulge components combined) to be $4.99^{+0.34}_{-0.50}\times 10^{10}$ 
M$_\odot$ in a $1.12\times10^{12}$ M$_\odot$ halo. Scaling down their result to 
coincide with our $10^{12}$ M$_\odot$ halo, the total mass of the galactic 
components in our simulation should be $M_{\mathrm{gal}} 
=4.46^{+0.30}_{-0.45}\times10^{10}$ M$_\odot$.

Since this measurement was obtained from stellar dynamics only,  it is 
sensitive to the presence of a mirror galactic component. However it is 
difficult to accurately estimate the contribution of ordinary baryons to 
the disk + bulge mass from the mass-luminosity relation. It is believed that 
about 20~\% 
of the baryons in the halo should condense into compact structures in the 
galaxy (see \cite{behroozi2010comprehensive} and references therein), which 
represents a visible 
matter contribution of $3.1\times 10^{10}$ M$_\odot$. This leaves room for
the remainder to come from a mirror galaxy component.\footnote{The fraction of condensed 
baryons fluctuates by a factor of $\sim 1.5$ from galaxy to galaxy, which is
consistent with  the disk + bulge components of our halo containing only 
ordinary baryons.}

Under this assumption, we derive a $2\,\sigma$ upper bound on the 
mass of the mirror galaxy (disk + bulge),
\be 
M_{\mathrm{gal}}' \lesssim 2\times10^{10}\ \mathrm{M}_\odot.
\label{mgalp}
\ee 

It is also possible to constrain the bulge mass of the MW separately.
Ref.\ 
\cite{cautun2019milky} determined $M_{\mathrm{bulge}}= 0.93^{+0.9}_{-0.8}\times 
10^{10}$ M$_\odot$ using Gaia DR2, in agreement with the value of ref.\  
\cite{sofue2013rotation}
obtained from rotation curves. A larger value was derived using photometric 
data from 
the VVV survey, estimating the contribution from visible stars to the bulge mass as 
$M_{\mathrm{bulge}}^{\mathrm{SM}} = 2.0\pm 0.3 \times10^{10}$ 
M$_\odot$ \cite{portail2015madetomeasure,zoccali2018weighing}.
Combining errors in quadrature, these imply the $2\,\sigma$ upper bound on the 
\MS\ contribution 
\be
 M_{\mathrm{bulge}}' =  M_{\mathrm{bulge}}-
 M_{\mathrm{bulge}}^{\mathrm{SM}}\lesssim 0.83\times10^{10}\ \mathrm{M}_\odot\,.
\label{mbulp}
\ee 

Both (\ref{mgalp}) and (\ref{mbulp}) imply limits comparable to that from 
the dark disk surface density, excluding
 $\beta\gtrsim1$ for any $x$. Due to the increased 
bulge fraction at large $x$, the bound on $M_{\mathrm{bulge}}'$ 
becomes tighter at large $x$, ruling 
out $\beta\gtrsim0.3$ at $x\simeq0.5$.

\subsection{Gravitational lensing}

Compact objects made of mirror matter could be detected through their 
gravitational lensing of distant stars, similar to more general ``MACHO'' models 
of DM. However, it is difficult to predict the microlensing rate from 
\MS\ structures since it  depends strongly on their masses. Like in the SM, these 
compact objects could include asteroids and comets, planets, molecular 
clouds, or stars and dense globular clusters, spanning over 15 orders of magnitude 
in mass. 
We will focus on compact objects of mass $10^{-1}$ M$_\odot\lesssim M 
\lesssim10$ M$_\odot$, corresponding to a main-sequence star or a small 
molecular cloud. As in the SM, smaller objects should represent a negligible 
fraction of the collapsed matter in the \MS.

Constraints on the MACHO fraction $f_{\mathrm{mac}}$ of DM in this mass range 
have been discrepant. The MACHO collaboration studied 
microlensing events towards the Large Magellanic Cloud (LMC) and initially
reported evidence that MACHOs of mass $(0.15-0.9)$ M$_\odot$ comprise 
(8-50)~\% of the total halo DM \cite{Alcock:2000ph}, but it was 
later 
found that their dataset was contaminated by variable stars 
\cite{Bennett:2005at}. 
The same survey showed no evidence for MACHOs in the mass range ($0.3-30$) M$_\odot$ 
\cite{Allsman:2000kg}. The EROS and OGLE surveys found no
evidence for MACHOs towards the LMC 
\cite{wyrzykowski2010ogle,wyrzykowski2009ogle,Tisserand:2006zx,1997A&A...324L..69R}
leading them to 
place an upper limit $f_{\mathrm{mac}}\lesssim (7-30)$~\%. 
The MEGA and POINT-AGAPE experiments came to different conclusions, 
the former finding no evidence for MACHOs towards M31 \cite{deJong:2005jm} 
while 
the latter reported $0.2\lesssim f_{\mathrm{mac}}\lesssim 0.9$ 
\cite{CalchiNovati:2005cd}.

To interpret these results we review some of the theory underlying MACHO searches.
Gravitational lensing is characterized by an optical depth
\be 
\tau =\frac{4\pi G D_s^2}{c^2} \int_0^{1} \rho(x)\, x (1-x)\, dx,
\ee  
where $D_s$ is the distance to the amplified star and the integral is taken 
along the line of sight, with $x$ in units of $D_s$. The optical depth is the instantaneous probability 
that a star's brightness is amplified by a factor of at least $1.34$, and 
is proportional to the mass 
density $\rho$ of the lens.

If $N_{\mathrm{s}}$ stars are monitored during a period $T_{\mathrm{obs}}$, 
then the expected number of detected microlensing events is
\be 
N_{\mathrm{ex}} = \frac{2}{\pi} 
\frac{T_{\mathrm{obs}}}{\langle 
t_E\rangle }\tau N_{\mathrm{s}}  \langle \epsilon \rangle,
\ee 
where $\langle t_E\rangle $ is the average Einstein radius crossing time and $ 
\langle \epsilon 
\rangle$ is an efficiency coefficient that depends on the experimental 
selection criteria.

All the constraints cited above assumed that the MACHOs have an isothermal 
density profile $\rho \sim (r^2+r_0^2)^{-1}$, which is often referred to as the 
``S model.''\ \  This assumption is not valid for 
mirror matter compact objects since they are preferentially distributed in the 
disk and the bulge of galaxies, like visible stars. Ref.\ \cite{Alcock:2000ph} 
estimated the total optical depth due to visible stars in the MW and the LMC 
galaxies as $\tau \simeq 2.4\times 10 ^{-8}$ with an average Einstein radius 
crossing time $\langle t_E \rangle \simeq 60$~days. 

The optical depth $\tau'$ due to a mirror galaxy is roughly proportional 
to its mass; we can therefore estimate it as
$\tau'\simeq\tau \beta \times (f_{\mathrm{mac}}/0.2)$, where $f_{\mathrm{mac}} = 
f_{\mathrm{disk}} + f_{\mathrm{bulge}}+f_{\mathrm{sat}}$ is the fraction of 
mirror particles that form compact objects in both the MW and its satellite 
galaxies. The factor of $0.2$ comes from the estimate that  
$\sim 20$~\% of the SM baryons in the halo end up in stars. In reality the 
contribution from each component weighs differently in the value of $\tau$: 
MACHOs in the LMC are about twice as likely to produce a lensing event as one 
located in the MW bulge or disk. To be more precise we should sum the optical 
depth $\tau_i' \simeq \tau_i (M_i'/M_i)$ of each component, where $M_i$ 
($M_i'$) is the mass of ordinary (mirror) stars in the LMC or in the MW bulge 
or disk. But the stellar masses of the LMC and of the individual MW components 
have large uncertainties and our simple treatment of satellite galaxies does not 
allow for an accurate identification of an LMC-like subhalo and the mass of its 
mirror galaxy. We can nevertheless make an order-of-magnitude estimate of $\tau'$ by 
putting all contributions on an equal footing and using the global fraction 
$f_{\mathrm{mac}}$ of condensed objects in the halo.

Since the Einstein radius is proportional to the square root of the mass of the 
lens \cite{Tisserand:2006zx}, the value of $\langle t_E \rangle$ can also be 
different in the \MS. 
Assuming a fiducial mass of $0.4\,M_\odot$ 
for SM stars, then we can approximate $\langle t_{E} ' \rangle = \langle 
t_{E} \rangle \sqrt{m_{\mathrm{mac}}/0.4\,M_\odot}$, where $m_{\mathrm{mac}}$ 
is the mirror MACHO mass.
 
The EROS-2 survey sets one of the most stringent limit on MACHOs in the direction
of the 
LMC. During $T_{\mathrm{obs}}=2500$~days, it monitored $N_{\mathrm{s}}= 
5.5\times 10^6$~stars and detected no microlensing event. This sets the 95~\% 
confidence limit $N_{\mathrm{ex}}<3$. From visible stars alone we expect 
$N_{\mathrm{ex}}\simeq 1.23$ events for an efficiency coefficient $\langle 
\epsilon\rangle \approx 0.35$. Then the limit on events from 
mirror stars is $N_{\mathrm{ex}}'\lesssim 
1.77$, giving
\be 
\beta f_{\mathrm{mac}} \lesssim 0.29 \left(\frac{0.35}{\langle \epsilon '
\rangle}\right) \sqrt{\frac{m_{\mathrm{mac}}}{0.4M_\odot}}\,,
\label{machoc}
\ee 
 where $\langle \epsilon '\rangle$ is the efficiency 
 coefficient of the \MS, which could differ from the SM value if the 
 MACHO mass is different. We will consider three benchmark values of 
 $m_\mathrm{mac}$ to constrain our model: 0.4~M$_\odot$, 1~M$_\odot$ and 
 10~M$_\odot$. For simplicity we will also assume $\langle \epsilon 
 '\rangle\approx0.35$ for all masses.
 
 The constraint (\ref{machoc}) is not very restrictive, despite mirror matter being
 capable of  forming roughly as many compact objects as visible matter.  If mirror
 stars  had a  mass  distribution similar to visible stars, then 
 $m_\mathrm{mac}\simeq0.4$~M$_\odot$ would only rule out $\beta\gtrsim2$, which  is
 already excluded by other observations.  It is possible that the  typical \MS\ MACHO
 mass exceeds that of SM stars since cooling and cloud  fragmentation are less efficient
 in the \MS, as we argued  in section  \ref{sec:starform}. In that case the bound would
 be relaxed even more. A full analysis of the stellar  evolution in the \MS, including
 heavier elements that we have not included, would be required to estimate 
 $m_{\mathrm{mac}}$ and the microlensing rate more accurately. But based on the 
present analysis, it seems unlikely  that MACHO
 detection  towards the LMC could be more constraining than 
 the disk surface density or the stellar mass in the MW.

\subsection{Bullet Cluster}
Interestingly, the Bullet Cluster allows us to set an upper limit on the hot 
gas fraction of mirror baryons, {\it i.e.,} the absence of structure formation in the 
\MS. 
The visible galaxies and stars on the scale of this cluster are essentially 
collisionless, but 
the hot gaseous baryons that surround the galaxies were impeded by dynamical 
friction and stripped from their hosts. Similarly, mirror galaxies and stars 
pass through each other unimpeded, just like CDM, while the hot 
clouds of mirror baryons will self-interact.

The most stringent constraint on DM comes from the survival of the smaller 
subcluster in the merger, as less than 30~\% of its mass inside a radius of 
150 kpc was stripped in the collision \cite{Markevitch:2003at}. This normally
yields a 
bound on the integrated cross section $\sigma/m$. Here we instead follow the approach 
of ref.\ \cite{Foot:2014mia}, constraining the distribution of mirror 
matter, in particular the mass of the hot gas fraction.  We recapitulate the argument as
follows.

Consider the elastic collision of two equal-mass mirror particles in the 
subcluster's reference frame. The incoming particles from the main cluster have 
an initial velocity $v_0 \approx4800$ km/s. After the 
collision, they scatter with velocities
\be 
v_1 = v_0 \cos \Theta, \quad v_2 =v_0\sin\Theta,
\ee
where $\Theta$ is the scattering angle of the incoming particle in the 
subcluster's frame.

For the subcluster to lose mass, both particles must be ejected from the halo: 
$v_1,v_2 > v_{\rm esc}$ where $v_{\rm esc}\approx 1200$ km/s is the escape velocity. 
This happens for a scattering angle $\theta$ (in the CM frame)
\be \label{eq:thetaLoss}
\frac{v_{\rm esc}}{v_0} < \sin\frac{\theta}{2} < 
\sqrt{1-\left(\frac{v_{\rm esc}}{v_0}\right)^2}.
\ee 
The scattering angles in the two frames are related by $\Theta = \theta/2$ for 
equal-mass particles. The evaporation rate is $R = N^{-1} dN/dt$ where $N$ is 
the 
total number of hot mirror particles in the subcluster. It can be expressed as
\cite{Kahlhoefer:2013dca}
\be 
R = n_2 v_0 \int_{\rm esc} \frac{d\sigma}{d\Omega_{CM}} d\Omega_{CM},
\label{evap}
\ee 
where $n_2$ is the number density of mirror particles in the main cluster and 
the bounds of the integral are given by eq.\ (\ref{eq:thetaLoss}).
Integrating (\ref{evap}) over the crossing time $t = w/v_0$, where $w$ 
is the width of the main cluster, leads to the fraction of evaporated hot 
mirror particles,
\be \label{eq:fracLost}
\frac{\Delta N}{N} = 1 - \exp\left(-\frac{\Sigma_2}{\mn}\int_{esc} 
\frac{d\sigma}{d\Omega_{CM}} d\Omega_{CM}\right),
\ee 
where $\Sigma_2$ is the surface density of the hot mirror matter gas in the 
main cluster. Taking the total DM surface density to be $\Sigma_{\mathrm{DM}}\simeq 
0.3$\,g/cm$^2$, we can estimate $\Sigma_2\simeq 
f_{\mathrm{gas}}^{\mathrm{BC}}(\Omega_{b'}/\Omega_{\mathrm{DM}}) 
\Sigma_{\mathrm{DM}}$, where $f_{\mathrm{gas}}^{\mathrm{BC}}$ is the hot mirror 
matter gas fraction in the main cluster. 

Because of the large mass of the cluster and the subcluster 
($M\gtrsim2\times10^{14}\,M_\odot$), the virial temperature of the mirror 
matter gas is high enough to fully ionize the H and He atoms. Mass 
evaporation therefore proceeds via Rutherford scattering between ions. 
Assuming that all mirror nuclei have a mass $\mn$ and a charge $Z=1+f\he$ (see 
eqs.\ (\ref{eq:fhe},\ref{eq:mn})), 
their differential cross section in the CM frame is
\be 
\frac{d\sigma}{d\Omega_{CM}}  = \left(\frac{ Z^2 \alpha}{4 E 
\sin^2(\theta/2)}\right)^2.
\ee 
where $E = \mn (v_0/2) ^2$ is the total kinetic energy in the CM frame. 
Plugging this in eq.\ (\ref{eq:fracLost}) and evaluating the integral within 
the bounds of eq.\ (\ref{eq:thetaLoss}) yields
\begin{align} 
\frac{\Delta N}{N} = 1 - &\exp\bigg\{\frac{-4\pi Z^4 \alpha^2 
\Sigma_2}{\mn^3 v_0^4}  \\ \nn
&\times\frac{1-2 \left(v_{esc}/v_0\right)^2}{ \left(v_{esc}/v_0\right)^2
\left(1- \left(v_{esc}/v_0\right)^2\right)} \bigg\}.
\end{align}

Assuming that only hot mirror particles are stripped in the 
collision, the constraint on the evaporated mass fraction of the subcluster is:
\be 
f_{\mathrm{evap}} = \frac{f_{\mathrm{gas}}^{\mathrm{BC}}\beta 
\Omega_b}{\Omega_{\mathrm{DM}}} 
\frac{\Delta N}{N} < 0.3.
\ee 

This does not apply directly to our study, since we specifically 
studied structure formation in a $10^{12}$ M$_\odot$ halo, 
while the Bullet subcluster has mass  $\sim 2\times10^{14}$ M$_\odot$. 
However ref.\ \cite{behroozi2010comprehensive} indicates that the stellar 
mass fraction in a Bullet subcluster-sized halo  is 
$\sim 10$~\% of the same fraction in a MW-like halo. We can therefore estimate 
the hot 
gas fraction of \MS\ matter in the Bullet Cluster as 
$f_{\mathrm{gas}}^{\mathrm{BC}}\simeq (1 - 0.1) f_{\mathrm{mac}}$ 
(recall that $f_{\mathrm{mac}}$ is the fraction of mirror matter compact objects 
in the central galaxy and its satellites, that we derived above).
However this is weaker than the kinematic data limits, and 
the resulting bound from the Bullet Cluster is similar in strength to
that from  microlensing, excluding only the region $\beta\gtrsim2$.

\subsection{Future constraints and signals}
\label{sec:future}
In this section we describe other astronomical observations that could 
lead to new 
constraints on the \MS\ in the next few years, as more data is collected and 
experimental sensitivity increases.

Gravitational wave (GW) astronomy is a promising new window to study our 
universe and the properties of DM. LIGO and other interferometer experiments 
are forecasted to put strong constraints on the fraction of primordial black holes 
(PBHs) in the universe, down to a mass scale of $\sim 10^{-13}$ M$_\odot$
\cite{Saito:2008jc,saito2009gravitationalwave,Carr:2009jm,Carr:2016drx}. 
However, the binary black hole (BBH) merger rate 
$\mathcal{R}_{BBH}^{exp}\sim9.7-101$ Gpc$^{-3}$ 
y$^{-1}$ detected by LIGO \cite{LIGOScientific:2018mvr} seems to exceed the 
predictions of $\mathcal{R}_{BBH}^{th}\sim5.4$ Gpc$^{-3}$ y$^{-1}$ 
in some theoretical models of star formation \cite{Askar:2016jwt}. 

In has been suggested in \cite{Beradze:2019dzc,Beradze:2019ujd} that this 
discrepancy could be explained by the early formation of BHs in mirror 
matter-dominated systems. This idea is supported by the fact that none of the 
GW signals from BBH mergers detected by LIGO were accompanied by an 
electromagnetic counterpart, indicating that those systems had accreted very 
little visible matter. A similar idea can be applied to binary neutron star 
(NS) mergers and BH-NS coalescence \cite{Beradze:2019yyp}, which only led to 
the detection of one electromagnetic signal \cite{Cowperthwaite:2017dyu} out of 
the many candidate events.

According to \cite{Beradze:2019dzc,Beradze:2019ujd}, since the cosmic star 
formation rate (SFR) peaked at $z\sim1.9$ for visible matter, then it should 
have peaked at a  redshift $z'\simeq-1+ (1+1.9)/x$ in the \MS, leaving more time
for mirror matter  to  form BHs and binary systems. According to our present
findings,  this argument is incorrect, since we have shown that star formation
depends  primarily on chemical abundances, matter temperature and the
gravitational  potential, not on the background radiation temperature. At late
times ($z\ll  z_{\rm dec}$),  visible and mirror particles collapse inside the 
same local gravitational potential well and they are shock-heated to the same
temperature  $\sim T\vir$ (recall eq.\ (\ref{eq:virTemp})). Hence the mirror
SFR  differs from that of the SM only because of its high He abundance and how
it impacts the  cooling rate. These effects are not encoded by a simple  
$x$-dependent rescaling of $z$. 

Nevertheless, the authors of \cite{DAmico:2017lqj,Latif:2018kqv} suggested that the 
inefficient cooling and fragmentation of mirror gas clouds could lead to the 
early formation of direct collapse black holes (DCBHs). Although they 
would more likely act as supermassive BH seeds, they could also increase the 
binary merger rate in the mass range probed by LIGO and the other GW 
interferometers. In the next decade, as the measurements and  
predictions for $\mathcal{R}_{BBH}$ are refined, as well as the 
understanding of BH formation from 
mirror matter, this could be a useful observable to further constrain such
models.

21-cm line surveys are another promising technique for studying late-time
cosmology and  structure formation. The EDGES experiment reported a surprisingly
deep  absorption feature in the signal emitted at the epoch of reionization 
\cite{Bowman:2018yin}. Although it still awaits confirmation, many have tried to
relate this anomaly to DM properties  \cite{Munoz:2018pzp, 
Fialkov:2018xre,Berlin:2018sjs,Barkana:2018cct,Liu:2019knx,Panci:2019zuu}. 
Mirror matter could be compatible with the EDGES result if 
the model is augmented by a large photon-mirror photon kinetic mixing term,
$\epsilon \sim 10^{-3}$, and if the CDM is light, $\sim 10$\,MeV. To
explain the EDGES anomaly would also
require breaking the mirror symmetry by allowing for a new long-range force
between the DM and the CDM, as shown in ref.\ \cite{Liu:2019knx}.  (The large
kinetic mixing would evade constraints from underground direct detection 
since millicharged mirror DM would not be able to penetrate the earth.) 
Ref.\ \cite{AristizabalSierra:2018emu} proposed an alternative mechanism in 
which mirror neutrinos decay to visible photons, $\nu_i'\to \gamma \nu_j$, to 
explain the EDGES anomaly, using a smaller kinetic mixing $\epsilon\lesssim 
10^{-6}$.  This scenario too would require mirror
symmetry breaking, in the form of a small \MS\ photon mass. These two models 
might require even further breaking of the mirror symmetry in order to avoid 
stringent 
limits $\epsilon\lesssim 10^{-9}-10^{-7}$ set by $N_{\rm eff}$ 
\cite{Foot:2014mia,Berezhiani:2008gi} and 
orthopositronium decay \cite{Vigo:2018xzc} in the unbroken symmetry 
scenario.  

Independently of whether the EDGES anomaly is confirmed,  furture 21-cm line 
surveys can be used to constrain compact DM objects like mirror stars. Should 
mirror matter compact objects form before visible stars (as in  
the early formation of DCBHs proposed by \cite{DAmico:2017lqj,Latif:2018kqv}), 
those objects would accrete visible matter and accelerate the reonization of 
the universe, leaving a characteristic imprint on the 21-cm signal
\cite{Mena:2019nhm} and distorting the CMB spectrum \cite{Ricotti:2007au}. The 
suppression of the power spectrum by a dark  sector, as we discussed in
section \ref{sec:Silk}, is also expected to delay structure  formation  and the
absorption feature of the 21-cm line \cite{Lopez-Honorez:2018ipk}.

It was recently suggested that gravitational lensing of fast radio bursts 
would present a characteristic interference pattern and could probe 
MACHOs in the mass range $10^{-4}$\,--\,$0.1$ M$_\odot$ \cite{Katz:2019qug}. 
Although this is smaller than the typical mass scale for mirror stars, it could 
lead to new constraints on the abundance of smaller objects,
 like mirror brown dwarfs and mirror planets. 

The idea that mirror planets could orbit visible stars (or the opposite) was 
proposed two decades ago \cite{Foot:1999ex,Foot:2000iu}, but not explored in 
detail.  A smoking gun signal for small mirror 
matter structures would be the detection of an exoplanet-like object via 
Doppler spectroscopy or microlensing without the expected transit, in the 
case where the inclination angle is $90^\circ$.
With improved understanding of how mirror planets form and 
how often they could be captured by a visible stars, the nondiscovery of such 
events could eventually rule out some of the parameter space of the model.

Finally, mirror stars  would heat and potentially dissolve visible wide binary 
star systems, star clusters and ultra-faint dwarf galaxies via dynamical 
relaxation. This effect was used to rule out heavy MACHOs ($m\gtrsim 5-10\, 
M_\odot$) from making up a significant fraction of DM 
\cite{Brandt:2016aco,2010ASPC..435..453Q}. Future studies of similar systems 
could tighten the constraints on MACHOs and, pending a more refined model for 
mirror star formation, on mirror matter.

\section{Early universe}
\label{sect:early}

One may wonder how likely it is to find an embedding of perfect mirror
symmetry in a complete model including inflation and baryogenesis, such that
the relative temperatures and baryon asymmetries in the two sectors differ as 
we have presumed.  These questions have been considered in earlier literature. 
Here we revisit them in light of more recent inflationary constraints.
 
\subsection{Temperature asymmetry}

A simple way of maintaining mirror symmetry while incorporating cosmological inflation
is to assume that each sector has its own inflaton, and to seek differences in their
reheating temperatures from initial conditions or other environmental effects, while
maintaining identical microphysics in each sector.
An early proposal for getting asymmetric reheating was given in 
ref.\ \cite{1985Natur.314..415K}, which proposed a `double-bubble inflation' 
model where the 
ordinary and mirror inflatons finish inflation by bubble nucleation
at different (random) times. In this case the 
first sector to undergo reheating gets exponentially redshifted until the 
second field nucleates a bubble of true vacuum.  However this is in the context
of ``old inflation'' driven by false vacua, which is untenable because the phase transitions
never complete.

A more promising mechanism was demonstrated in  
ref.\ \cite{Berezinsky:1999az}, which considered
two-field chaotic inflation with decoupled quadratic potentials, with total potential of the
form
\be
	V_{\rm tot} = V(\phi) + V(\phi')
\ee
plus respective couplings of each field to its own sector's matter particles, 
to accomplish reheating.  It was shown that the solutions are such that 
the ratio of the two inflatons $\phi'/\phi$ remains constant
during inflation. Then the ratio of the reheating temperatures goes as 
$(\phi'/\phi)^{2/3}$, and is thereby determined
by the random initial conditions.  This idea is now ruled out by Planck data \cite{Akrami:2018odb},
strongly disfavoring chaotic inflation models, that have concave potentials. 

We suggest a possible way of saving this scenario; one can flatten the
potentials at large field values  using nonminimal kinetic terms
\cite{Lee:2014spa}, for example
\be
	{\cal L}_{\rm kin} = \left(1+ f {\phi^4\over m_P^4}\right) (\partial\phi)^2 + 
\left(1+ f {\phi'^4\over m_P^4}\right) (\partial\phi')^2
\ee
For $\phi,\phi'\gg m_P$, the canonically normalized fields are $\chi \sim \phi^3$,
$\chi'\sim \phi'^3$, so that a potential of the form $m^2(\phi^2+\phi'^2)$ 
becomes proportional to $(\chi^{2/3} + \chi'^{2/3})$, mariginally consistent with Planck
constraints on the tensor-to-scalar ratio and spectral index, while maintaining the
separability of the potential.  In common with the simpler model, the trajectory in field space 
is a straight line towards the vacuum at $\phi=\phi'=0$ and the initial 
conditions determine the ratio of reheat temperatures, $x=T_R'/T_R$.  We leave this
for future investigation.

Alternatively, one could imagine there is just a single inflaton, that is 
charged under the mirror symmetry such that $\phi \to -\phi$, and couples to the
Higgs fields of the two sectors with opposite signs,
\be \label{eq:preheat}
V \supset V_{\rm inf}(\phi) + \frac{\mu}{2} \phi
\left(h'^2-h^2\right) + V(h) + V(h')
\ee  
so as to preserve the mirror symmetry.
At the end of inflation, $\phi$  oscillates about its minimum at $\phi = 0$, 
resulting in a time-dependent frequency $\omega_k$ for the Fourier modes of the 
Higgs fields \cite{Dufaux:2006ee},
\be 
\omega_k ^2 =  m^2_h +\frac{k^2}{a^2} \pm \frac{\mu \Phi}{a^{3/2}} \sin (m_\phi 
t),
\ee 
where $\Phi$ is the amplitude of the inflaton at the beginning of preheating. 
Both fields are periodically tachyonic
whenever $\omega_k ^2<0$, resulting in an exponential growth 
of the occupation number: $n_k\sim \prod_{j} \exp(X_k^j)$, where 
$X^j_k$ is the 
particle production rate during the $j$th inflaton oscillation.\footnote{When 
$\omega_k^2>0$, the modes  also grow via parametric resonance, but 
tachyonic resonance is known to be a much more efficient preheating 
mechanism \cite{Dufaux:2006ee,Abolhasani:2009nb}.}\ \  

With the expansion of the universe, the particle production efficiency 
decreases: 
$X_k^j \sim a(t_j)^{-3/4}$ \cite{Abolhasani:2009nb}, where $a(t_j)$ is the 
average scale factor during the 
$j$th tachyonic phase. Because of their opposite coupling to the inflaton, the 
tachyonic resonances of $h_k$ and $h'_k$ are out of phase with each other, 
resulting in different growth rates. In particular, the 
first field to experience tachyonic instability gives rise to dominant reheating 
into its own sector, creating a temperature difference between the two.
Details will be given in a future publication.

\subsection{Baryogenesis}

A further challenge is to explain how the \MS\ baryon asymmetry could attain values compatible
with the allowed regions from our analysis.  
The baryon densities in the two sectors are related by
\be
	\eta_{b'} = {\beta\over x^3}\eta_{b}\,.
\ee
where $\eta_{b'} = n_{b'}/s'$, with $s'$ being the entropy density of the \MS, 
while $\eta_{b} = n_{b}/s$, where $s$ is the SM entropy density.
Hence for $\beta\sim 0.3$ and $x \sim 0.5$, for example, we require would baryogenesis
in the \MS\ to be more efficient than in the SM: $\eta_{b'}/\eta_{b}=2.4$. 
Lowering the value of $x$ with $\beta$ fixed requires an even greater 
efficiency for mirror baryogenesis.

Leptogenesis may offer a viable explanation for this mild hierarchy.  Naively 
one could 
expect that $\beta~\simeq~x^3$ since this is the ratio of the densities of 
the decaying heavy neutrinos in the two sectors, which would lead to a very 
small mirror baryon abundance. In particular, if $x\lesssim0.1$ mirror 
particles would be too sparse to produce any observable signal. However this 
assumes that the washout factor is the same in both sectors, which need not
be the case.  At any given time, mirror 
heavy neutrino decays occur farther out of equilibrium than those of their
SM counterparts, due to their lower temperature (while the Hubble rate is the 
same for both
sectors).  This can make the washout factor smaller in the \MS\ 
\cite{Buchmuller:2004nz}
leading to more efficient leptogenesis in that sector.  This conclusion is 
compatible with
refs.\ \cite{1985Natur.314..415K,Berezhiani:2000gw}, which considered the 
analogous mechanism of GUT baryogenesis, that
also relies upon an asymmetry produced by out-of-equilibrium decays.  They also 
found
that $\beta \sim x^3$ for particles decaying well out of equilibrium, while 
$\beta\sim x$ in the
strong washout regime.  Hence for some intermediate choice it should be 
possible to have
$x^3 < \beta < x$ as in our example of $\beta \sim 0.3$ and $x\sim0.5$. In 
general these scenarios predict that baryogenesis in the \MS\ is at least as 
efficient as in the SM, $\eta_{b'}/\eta_{b}\geq1$.

Ref.\ \cite{Berezhiani:2000gw} in addition considered electroweak baryogenesis 
in the \MS,
and finds that $\beta =x^3$ is predicted for the typical case in which the 
phase transition
is strong enough so that sphalerons are highly suppressed inside the bubbles of 
true
electroweak vacuum.  For finely tuned scenarios in which sphalerons washout is 
important for
attaining the final baryon asymmetry, larger values of $\beta$ could achieved, 
since the
phase transitions happens earlier in the SM sector, and thus gives longer time 
for sphalerons
to washout the initial asymmetry.

\subsection{Non-minimal mirror matter model}

Although it goes beyond the scope of the present investigation, it is
interesting to contemplate less minimal scenarios in which mirror
symmetry is not exactly conserved at the microscopic level.  This of course
makes it easier to achieve the asymmetry between temperatures of the two
sectors.

A simple example is to allow for the mirror Higgs field to have a different VEV, 
$v'\neq v$, which changes the mirror fermions masses by the factor $v'/v$. If 
$v'/v>1$ and we introduce portal interactions between the two 
sectors in the early universe, there would be a net transfer of entropy to 
the less massive SM fermions until the two sectors decouple 
from each other and their temperature ratio freezes out. If mirror symmetry is 
already broken during reheating, it could affect the decay rate of the inflaton 
into each sector, leading to different reheating temperatures without 
portal interactions \cite{Berezhiani:1995am}.
Other mechanisms for the broken mirror parity 
scenario can be adapted from similar theories like twin Higgs models (see 
\cite{Chacko:2018vss} and references therein).
By generalizing the chemical and cooling rates described in 
Appendices \ref{appA} and \ref{appB} for the nonsymmetric \MS, our 
present analysis could be repeated to study structure formation in this altered
scenario.

Another variation of the model, already alluded to in 
section  \ref{sec:future} is the inclusion of the Higgs portal 
interaction $h^2 h'^2$ or kinetic mixing $F^{\mu\nu}F'_{\mu\nu}$ 
between the \MS\
and the SM.  Although they are significantly
constrained by laboratory and astrophysical considerations,  they
could still have important implications for cosmology and structure 
formation. In particular, ref.\ \cite{Foot:2014mia} argued that 
mirror photons produced in ordinary supernovae would heat the dark
\MS\ disk,  leading to its expansion. Conversely, visible photons
could be produced in  early mirror supernovae and accelerate the
reionization of ordinary baryons  \cite{Foot:2004kd}. These portal
interactions would also open the possibility  for direct detection
experiments and give characteristic astronomical signals 
\cite{Curtin:2019ngc}.

\section{Conclusions} \label{sect:conc} 

Working within the context of unbroken mirror symmetry, we have investigated the formation of
dark galactic structures of  mirror matter in a MW-like halo and constrained the parameters
$x=T'/T$, $\beta=\Omega_{b'}/\Omega_b$ of the  theory using astrophysical data.   By 
our assumption, all chemical and nuclear processes  have the same rates in
each sector, but mirror baryons turn out to be He-dominated because of their lower temperature.

The lower temperature and large He abundance of the \MS\ have many consequences for
its cosmology and structure formation. H and He recombination are generally more 
efficient, leaving a lower density of free electrons at late times. While H$_2$ 
formation is also more efficient in the \MS, its residual density is suppressed 
by the low H abundance. H$_2$ and free electrons are the two main cooling 
channels of hot gas clouds in the SM; their low abundances in the \MS\ imply 
that cooling and fragmentation of mirror gas clouds are less efficient, which 
alters structure formation. 
We find that primordial mirror stars are much more massive than their visible  counterparts.
Because of the He fraction, the lifetime of  mirror stars is drastically shortened, which in
turns increases the SN feedback on collapsing gas clouds in the \MS.

Overall, the formation of mirror galaxies is strongly inhibited for 
$\beta\lesssim 0.5$ and $x\lesssim0.1$. For such parameters, 
mirror baryons tend to stay in an isothermal hot gas cloud, avoiding 
constraints from astronomical and cosmological data, making such a scenario
difficult to distinguish from more generic models containing a subdominant component of 
dissipative dark matter.
The most stringent such constraints come from observations of the MW disk surface 
density and bulge mass, which rule out $\beta\gtrsim 0.3$ at $x=0.5$ and 
$\beta\gtrsim0.8$ for $x\lesssim 0.1$. Both of these are derived by 
comparing stellar kinematics (measured \textit{e.g.} by  Gaia) with 
spectroscopy data. One can therefore hope that the release of Gaia EDR3 in 2020 
and improved understanding of luminosity data will 
shed more light on the existence of dark galactic structures. 21-cm line 
surveys and gravitational wave astronomy are also promising leads to explore 
the properties of DM.

It is theoretically challenging to generate a temperature asymmetry between the mirror and
visible  particles  while maintaining unbroken mirror symmetry at the microscopic level. We
proposed several   ideas that could give rise to asymmetric reheating, which intuitively are
expected to produce only a small hierarchy with $x\gtrsim 0.1$.  Coincidentally this is the
most constrained region of the model, hence the most interesting from the perspective of
discovery.

\smallskip
{\bf Acknowledgment.} We thank A.\ Benson, A.\ Ghalsasi, M.\ 
McQuinn 
and A.\ Lupi for very helpful correspondence. Our work is supported by NSERC 
(Natural Sciences and Engineering Research Council, Canada).  JSR is also 
supported by the FRQNT (Fonds de recherche du Qu\'ebec -- Nature et 
technologies, Canada).\\

\begin{appendix}

\section{Recombination evolution equations}
\label{appA}

Here we define quantities appearing in the evolution equations (\ref{eq:recHii}-\ref{eq:recTm}) that are needed
for recombination in the mirror sector.

The Thomson
scattering cross section is $\sigma_T$ is and $a_R = 4\sigma/c$ is the radiation constant,
related to the Stefan-Boltzmann constant $\sigma$. 
The other parameters come from the atomic configuration of both elements. 
The H$^0$ 2s-1s frequency is $\nu\h= 2466.0$~THz, the He$^0$ 2$^1$s-1$^1$s 
frequency
is $\nu\he =4984.9$~THz and $\tilde{\nu}\he = 145.62$~THz is the frequency 
difference between the 2$^1$p-1$^1$s and the 2$^1$s-1$^1$s transitions of 
He$^0$.
The two-photon rates are $\Lambda\h = 8.22458$~s$^{-1}$ and $\Lambda\he =
51.3$~s$^{-1}$.

The two recombination parameters $\alpha_i$ are given by (in m$^3$ s$^{-1}$):
\begin{gather}
	\alpha\h = \frac{F}{10^{19}} \frac{a t^b}{1+c t^d},\label{eq:alphaH}\\
	\alpha\he = q \left[ \sqrt{\frac{T_M'}{T_2}}\left(1+\sqrt{\frac{T_M'}{T_2}}\right)^{1-p}\left(1+\sqrt{\frac{T_M'}{T_1}}\right)^{1+p}\right]^{-1} ,\label{eq:alphaHe}
\end{gather}
where the fit coefficients are $a = 4.309$, $b =-0.6166$, $c=0.6703$, $d=0.5300$
and $t=T_M'/10^4$~K. $F$ is a fudge factor set to 1.125 \cite{Giesen:2012rp}. 
Furthermore, $q=10^{-16.744}$, $p = 0.711$, $T_1 = 10^{5.114}$~K and $T_2$ was 
fixed at 3~K. The principle of detailed balance gives the photoionization 
coefficients $\beta_i$:
\be  \label{eq:betaRec}
\beta_i = g_i\alpha_i \left(\frac{m_e kT_M'}{2\pi \hbar^2}\right)^{3/2} 
e^{-\chi_i/kT_M'}.
\ee 
The statistical weight factor $g_i$ is 1 for H and 4 for He and the ionization
energies from the 2s level are $\chi\h = 3.3996$~eV and $\chi\he= 3.9716$~eV.

Finally, the coefficients $K_i$ take into account the cosmological redshift of 
the H Ly$\alpha$ and He$^0$ 2$^1$p-1$^1$s photons that reionize the atoms. They 
are given by $K_i = \lambda_i^3/(8\pi H(z))$ with $\lambda\h = 121.5682$~nm and 
$\lambda\he =58.4334$~nm.

\section{Cooling rates and chemical abundances}
\label{appB}
Here we describe the various processes that contribute to the cooling of
dark baryons, and their rates.

(i) Inverse Compton scattering.
At early times, electrons can exchange energy with the background photons,
with cooling rate
\be 
\cool_{\mathrm{Comp}} = \frac{4 T_M}{m_e} \sigma_T n_e a_R T^4.
\ee 
where $T_M$ is the matter temperature (given by eq.\ \ref{eq:virTemp}) and 
$T$ is the radiation temperature. Because the expansion of the universe 
redshifts $T$, inverse Compton cooling becomes negligible at late times.

(ii) Brehmsstrahlung.  At very high temperatures the gas will be fully ionized and will primarily cool 
via free-free emissions (bremsstrahlung), whose cooling rate is:
\be 
\cool_{\mathrm{ff}} = \frac{16 \alpha^3 g_{\mathrm{ff}}}{3} \left(\frac{2 \pi 
T}{3m_e^3}\right)^{1/2} 
n_e \sum_{\mathrm{ions}} n_i Z_i^2.
\ee 
The sum runs over the ionized species (H$^+$, He$^+$ and He$^{++}$) and $Z_i$ 
is their electric charge. For our analysis we took the Gaunt factor to be 
$g_{\mathrm{ff}}\simeq 1$.

(iii) Atomic transitions. When the ionization fraction of the gas is too small, bremsstrahlung becomes 
inefficient. At this point atomic processes take the lead in the cooling of the 
gas. As ions and free electrons recombine to form neutral atoms, they radiate 
energy. Atoms can also collide with free electrons which will temporarily 
excite or ionize the atom until they return to their ground state. The atomic 
cooling rates $\cool_{\mathrm{atom}}$ are given in table \ref{tab:atomCool}.

(iv) Molecular transitions.  Atomic cooling can only bring the gas to a temperature of $\sim10\,000$ K 
(about 1 eV), since below this point electrons don't carry enough energy to 
excite or ionize the atoms. But unlike atoms, molecular hydrogen possesses 
rotational and vibrational modes which are easily excited by collisions. 
As the molecules return to their ground state, they emit low-energy photons 
which allow the temperature to drop to $\sim 200$ K if H$_2$ is sufficiently 
abundant.

The cooling function for molecular hydrogen can be parametrized as follows 
\cite{1979ApJS...41..555H, Grassi:2013lha,Glover:2008pz}:
\be 
\cool_{\mathrm{mol}} =\frac{n\hh L_{\mathrm{LTE}}}{1+ 
L_{\mathrm{LTE}}/L_{\mathrm{low}}}
\ee 
The $L$'s are cooling coefficients associated with rotational and vibrational 
modes excited by collisions with other species, either in local thermodynamic 
equilibrium (LTE) or in the low density regime. We can split the LTE 
coefficient into the contributions from rotational and vibrational excitations:
$L_{\mathrm{LTE}} = L_{\mathrm{LTE}}^{\mathrm{rot}} + 
L_{\mathrm{LTE}}^{\mathrm{vib}}$ \cite{1979ApJS...41..555H}, where:
\begin{align}
	\begin{split}
	L_{\mathrm{LTE}}^{\mathrm{rot}}=&\bigg[\left(\frac{9.5\times10^{-22} 
		T_3^{3.76}}{1+0.12 T_3^{2.1}}\right) e^{-(0.13/T_3)^3} \\ 
		&\qquad+ 	(	3\times10^{-24}) e^{-0.51/T_3}\bigg] \mathrm{~~erg\ 
		s^{-1}}
	\end{split}\\
\begin{split}
	L_{\mathrm{LTE}}^{\mathrm{vib}} =& \bigg[ (6.7\times10^{-19}) 
	e^{-5.86/T_3}
	\\ &\qquad + 
1.6\times10^{-18}e^{-11.7/T_3}\bigg]  \mathrm{~~erg\ s^{-1}}
\end{split}
\end{align}
In these expressions $T_3 = T/(10^3\ \mathrm{K})$.

In the low density limits, each species excite H$_2$ with a different rate. 
Therefore we can write 
\be
L_{\mathrm{low}} = \sum_k L_k n_k,
\ee
where $k$ represents either H$^0$, H$^+$, H$_2$, He or $e$ and the $L_k$ are 
determined from a fit of the following form:
\be 
\log_{10} L_k = \sum_{i=0}^{N} a_i^{(k)}\, \log_{10}\! T_3.
\ee 
All fit coefficients $a_i$ are given in table \ref{tab:h2Cool}. 

At late times the intensity of the photon background is negligible, which is 
why we only considered ionization and excitation from collisions with matter 
and not with background photons. Also note that all cooling rates given above 
are valid as long as the gas is optically thin. If the density is too high, the 
emitted photons can't escape the gas and the energy loss is slowed down. In 
this approximation we can also ignore any heating process that would counter 
the cooling.

To compute those cooling rates, one must also specify the density of each 
chemical species. In the steady-state approximation the densities are given by 
eq.\ (\ref{eq:steady}) where are the necessary rates $k_i$ are listed in table 
\ref{tab:rates}.

\begin{table*}[t] 
	\centering
	\caption{Cooling rates for atomic processes. $T_K$ is 
		the 
		gas temperature in 
		kelvin and $T_n=T/(10^n\mathrm{\ K})$. The densities $n_i$ are in 
		cm$^{-3}$. Adapted from \cite{Cen:1992zk,MBW:2010}.}\label{tab:atomCool}
	\begin{tabular}{ccc}
		\hline\hline 
		Process & Species & $\cool_{\mathrm{atom}}$ (erg s$^{-1}$ cm$^{-3}$)\\ 
		\hline
		
		\multirow{3}{*}{\shortstack{Collisional\\ excitation}} & H$^0$ & 
		$7.5\times10^{-19} (1+T_5^{1/2})^{-1} e^{-118348/T_K} n_e n\hi$ \\ 
		& He$^+$ & $5.54\times10^{-17} T_K^{-0.397}(1+T_5^{1/2})^{-1} 
		e^{-473638/T_K} n_e n\heii$\\
		& He$^0$ (triplets) & $9.10\times10^{-27} 
		T_K^{-0.1687}(1+T_5^{1/2})^{-1} 
		e^{-13179/T_K} n_e^2 	n\heii$\\[6pt] 
		
		\multirow{3}{*}{\shortstack{Collisional\\ ionization}} & H$^0$ & 
		$1.27\times10^{-21} T_K^{1/2}(1+T_5^{1/2})^{-1} 
		e^{-157809.1/T_K} n_e n\hi$	\\
		& He$^0$ & $9.38\times10^{-22} T_K^{1/2}(1+T_5^{1/2})^{-1} 
		e^{-285335.41/T_K} n_e n\hei$\\
		& He$^+$ & $4.95\times10^{-22} T_K^{1/2}(1+T_5^{1/2})^{-1} 
		e^{-631515/T_K} n_e n\heii$\\ 
		& He$^0$($2^3S$) & $5.01\times10^{-27} T_K^{-0.1687}(1+T_5^{1/2})^{-1} 
		e^{-55338/T_K} n_e^2 n\heii$\\ [6pt]
		\multirow{3}{*}{Recombination} & H$^+$ & 
		$8.7\times10^{-27} T^{1/2} T_3^{-0.2} (1+T_6^{0.7})^{-1} n_e n\hii$\\
		& He$^+$ & $1.55\times10^{-26} T^{0.3647} n_e n\heii$\\
		& He$^{++}$ &$3.48\times10^{-26} T^{1/2} T_3^{-0.2} (1+T_6^{0.7})^{-1} 
		n_e 
		n\heiii$ \\[6pt]
		\shortstack{Dielectronic\\recombination} & He$^+$ &
		$1.24\times10^{-13} T^{-1.5}e^{-470000/T_K}(1+0.3 e^{-94000/T_K}) n_e 
		n\heii$\\
		\hline
	\end{tabular}

\end{table*}

\renewcommand{\arraystretch}{1}

\begin{table*}
	\centering
	\caption{Fitting coefficients for H$_2$ cooling rates in the low density 
		limit assuming a 3:1 ortho-para ratio. Adapted from 
		\cite{Grassi:2013lha}.}\label{tab:h2Cool}
	\begin{tabular}{cclccl}
		\hline\hline
		Species\ &\ Temperature range (K)\ &\ Coefficients\ &\ Species\ &\  
		Temperature
		range (K)\ &\ Coefficients\ \\
		\hline
		H$^0$ & $10 < T \le 100$ & $a_{0} = -16.818342$ & H$^0$ & $100 < T \le 
		1000$ & 
		$a_{0} = -24.311209$ \\
		& &  $a_{1} =  37.383713 $ & & & $a_{1} = 3.5692468$ \\
		& &   $a_{2} = 58.145166 $ & & & $a_{2} =  -11.332860$ \\
		& &   $a_{3} = 48.656103 $ & & & $a_{3} = -27.850082$ \\
		& &   $a_{4} = 20.159831 $ & & & $a_{4} = -21.328264$ \\
		& &   $a_{5} = 3.8479610 $ & & & $a_{5} = -4.2519023$ \\
		& &  & && \\
		%\hline
		H$^0$ & $1000 < T \le 6000$ &  $a_{0} = -24.311209$ & H$_2$ & $100 < T 
		\le 
		6000$ & $ a_{0} =  -23.962112$ \\
		& & $a_{1} = 4.6450521$   &  & & $a_{1} = 2.09433740$ \\
		& & $a_{2} =  -3.7209846$ &  & & $a_{2} =  -0.77151436$ \\
		& & $a_{3} = 5.9369081$   &  & & $a_{3} = 0.43693353$ \\
		& & $a_{4} = -5.5108047$  &  & & $a_{4} = -0.14913216$ \\
		& & $a_{5} =  1.5538288$  &  & & $a_{5} = -0.033638326$ \\
		& & & & & \\
		%\hline
		He$^0$ & $10 < T \le 6000$ & $a_{0} = -23.689237$ &  H$^+$ & $10 < T 
		\le 
		10000$ & $a_{0} = -21.716699$ \\
		& & $a_{1} =   2.1892372$  &  & & $a_{1} =  1.3865783$ \\
		& & $a_{2} =  -0.81520438$ &  & & $a_{2} =  -0.37915285$ \\
		& & $a_{3} =  0.29036281$  &  & & $a_{3} =  0.11453688$ \\
		& & $a_{4} =  -0.16596184$ &  & & $a_{4} = -0.23214154$ \\
		& & $a_{5} =  0.19191375$  &  & & $a_{5} = 0.058538864$ \\
		& & && & \\
		%\hline
		$e$ & $10 < T \le 200$ & $a_{0} =  -34.286155$ &  $e$ & $200 < T 
		\le 10000$ & $a_{0} = -22.190316$ \\
		& & $a_{1} = -48.537163 $  & & & $a_{1} = 1.5728955$ \\
		& & $a_{2} =  -77.121176$  & & & $a_{2} = -0.21335100$ \\
		& & $a_{3} =  -51.352459$  & & & $a_{3} = 0.96149759$ \\
		& & $a_{4} =  -15.169160 $ & & & $a_{4} = -0.91023195$ \\
		& & $a_{5} = -0.98120322$  & & & $a_{5} = 0.13749749$ \\
		%\hline
		\hline
	\end{tabular}
\end{table*}

\begin{table*}
	\caption{Chemical reaction rates considered in our analysis. $T_K$ and 
		$T_e$ represent the gas temperature in K and eV, respectively, while 
		$T_{\gamma,e}$ is the photon temperature in eV. Table 
		adapted from \cite{Grassi:2013lha, Hirata:2006bt, Abel:1996kh}. Some 
		minor 
		reactions were ignored for simplicity.}\label{tab:rates}
	\begin{tabular}{llc}
		\hline\hline
		Reaction & Rate coefficient (cm$^3$ s$^{-1}$ or s$^{-1}$) & Temperature 
		range \\	
		\hline
		1) H$^0$ + $e$ $\rightarrow$ H$^+$ + 2$e$  & $k_1$ = 
		exp[-32.71396786+13.5365560 ln $T_e$ &   \\
		& - 5.73932875 (ln $T_e$)$^2$+1.56315498 (ln $T_e$)$^3$ &   \\
		& - 0.28770560 (ln $T_e$)$^4$+3.48255977 $\times$ 10$^{-2}$(ln 
		$T_e$)$^5$ &  \\
		& - 2.63197617 $\times$ 10$^{-3}$(ln $T_e$)$^6$+1.11954395 $\times$ 
		10$^{-4}$(ln $T_e$)$^7$ & \\
		& - 2.03914985 $\times$ 10$^{-6}$(ln $T_e$)$^8$] & \\
		2) H$^+$ + $e$ $\rightarrow$ H$^0$  + $\gamma$ & $k_2$ = 3.92 $\times$ 
		10$^{-13}$ $T_e$ $^{-0.6353}$ & $T \le 5500$ K  \\
		& $k_2$ = $\exp$[-28.61303380689232 & $T > 5500$ K  \\
		& - 7.241 125 657 826 851 $\times$ 10$^{-1}$ ln $T_e$&\\ 
		& - 2.026 044 731 984 691 $\times$ 10$^{-2}$ (ln $T_e$)$^2$&\\
		& - 2.380 861 877 349 834 $\times$ 10$^{-3}$ (ln $T_e$)$^3$&\\
		& - 3.212 605 213 188 796 $\times$ 10$^{-4}$ (ln $T_e$)$^4$&\\
		& - 1.421 502 914 054 107 $\times$ 10$^{-5}$ (ln $T_e$)$^5$&\\
		& + 4.989 108 920 299 510  $\times$ 10$^{-6}$ (ln $T_e$)$^6$&\\
		& + 5.755 614 137 575 750  $\times$ 10$^{-7}$ (ln $T_e$)$^7$&\\
		& - 1.856 767 039 775 260  $\times$ 10$^{-8}$  (ln $T_e$)$^8$&\\
		& - 3.071 135 243 196 590  $\times$ 10$^{-9}$  (ln $T_e$)$^9$] & \\
		3) He$^0$ + $e$ $\rightarrow$ He$^+$ + 2$e$  & $k_3$ = 
		$\exp$[-44.09864886 
		&  \\
		& + 23.915 965 63 ln$T_e$& \\
		& - 10.753 230 2 (ln $T_e$)$^2$&\\
		& + 3.058 038 75 (ln $T_e$)$^3$&\\
		& - 5.685 118 9 $\times$ 10$^{-1}$ (ln $T_e$)$^4$&\\
		& + 6.795 391 23 $\times$ 10$^{-2}$ (ln $T_e$)$^5$&\\
		& - 5.009 056 10 $\times$ 10$^{-3}$ (ln $T_e$)$^6$&\\
		& + 2.067 236 16 $\times$ 10$^{-4}$ (ln$T_e$)$^7$&\\
		& - 3.649 161 41 $\times$ 10$^{-6}$ (ln $T_e$)$^8$] &  \\
		4) He$^+$ + $e$ $\rightarrow$ He$^0 $ + $\gamma$ & $k_4$ =  3.92 
		$\times$ 
		10$^{-13}$ $T_e$ $^{-0.6353}$ &  $T_e \le 0.8$  \\
		& $k_4 = $  3.92 $\times$ 10$^{-13}$ $T_e^{-0.6353}$ & $T_e > 0.8$  \\
		& + 1.54 $\times$ 10$^{-9}$ $T_e^{-1.5}$ [1.0 + 0.3 / $\exp$(8.099 328 
		789 667/$T_e$)] & \\
		& /[$\exp$(40.496 643 948 336 62/$T_e$)]&\\
		&  & \\
		5) He$^+$ + $e$ $\rightarrow$ He$^{++}$ + 2$e$ & $k_5$ = 
		$\exp$[-68.710 409 902 120 01 & \\
		& + 43.933 476 326 35 ln$T_e$& \\
		& - 18.480 669 935 68 (ln $T_e$)$^2$& \\
		& + 4.701 626 486 759 002 (ln $T_e$)$^3$ &\\
		& - 7.692 466 334 492 $\times$ 10$^{-1}$ (ln $T_e$)$^4$&\\
		& + 8.113 042 097 303 $\times$ 10$^{-2}$ (ln $T_e$)$^5$&\\
		& - 5.324 020 628 287 001 $\times$ 10$^{-3}$ (ln $T_e$)$^6$&\\
		& + 1.975 705 312 221 $\times$ 10$^{-4}$ (ln $T_e$)$^7$& \\
		& - 3.165581065665 $\times$ 10$^{-6}$ (ln $T_e$)$^8$] & \\ 
		6) He$^{++}$ + $e$ $\rightarrow$ He$^+$ + $\gamma$ & $k_6$ = 3.36 
		$\times$ 10$^{-10}$ $T_K^{-1/2} (T_K/1000)^{-0.2} 
		(1+(T/10^6)^{0.7})^{-1}$ 
		& \\ \hline
	\end{tabular}
\end{table*}

\addtocounter{table}{-1} %the previous table continues here
\begin{table*}
	
	\caption{\textit{(Continued)} Chemical reaction rates considered in our 
		analysis. $T_K$ and 
		$T_e$ represent the gas temperature in K and eV, respectively, while 
		$T_{\gamma,e}$ is the photon temperature in eV. Table 
		adapted from \cite{Grassi:2013lha, Hirata:2006bt, Abel:1996kh}. Some 
		minor 
		reactions were ignored for simplicity.}
	\begin{tabular}{llc}
		\hline\hline
		Reaction & Rate coefficient (cm$^3$ s$^{-1}$  or s$^{-1}$) & 
		Temperature range\\	
		\hline
		7) H$^0$ + $e$ $\rightarrow$ H$^-$ + $\gamma$ & $k_7$ = 3 $\times$ 
		10$^{-16}$ 	$(T_K/300)^{0.95}$ $\exp(-T_K/9320)$ & \\
		-7) H$^{-}$ + $\gamma$ $\rightarrow$ H$^0$ + $e$ & 	$k_{-7} =4\ k_7 
		\left({m_e T_{\gamma,e}}/{2\pi\hbar^2}\right)^{3/2} 
		\exp(-0.754/T_{\gamma,e})$ & \\

		8) H$^-$ + H$^0$ $\rightarrow$ H$_2$ + $e$ & $k_8 = 1.5\times 10^{-9} 
		\left({T_K}/{300}\right)^{-0.1}$ & \\
		
		11) H$_2$ + H$^+$ $\rightarrow$ H$_2^+$ + H$^0$\ \ & $k_{11}$ = 
		$\exp$[-24.249 
		146 877 315 36 & \\
		& + 3.400 824 447 095 291 ln $T_e$& \\
		& - 3.898 003 964 650 152  (ln$T_e$)$^2$&\\
		&+ 2.045 587 822 403 071 (ln $T_e$)$^3$&\\	
		& - 5.416 182 856 220 388 $\times$ 10$^{-1}$ (ln $T_e$)$^4$&\\
		& + 8.410 775 037 634 12 $\times$ 10$^{-2}$ (ln $T_e$)$^5$&\\
		& - 7.879 026 154 483 455 $\times$ 10$^{-3}$ (ln $T_e$)$^6$&\\
		& + 4.138 398 421 504 563 $\times$ 10$^{-4}$ (ln $T_e$)$^7$&\\
		& - 9.363 458 889 286 11 $\times$ 10$^{-6}$ (ln $T_e$)$^8$] & \\
		
		12) H$_2$ + $e$ $\rightarrow$ 2H$^0$ + $e$ & $k_{12}$ = 5.6 $\times$ 
		10$^{-11} T_K^{0.5} \exp(-102124.0/T_K)$ & \\
		13) H$^-$ + $e$ $\rightarrow$ H$^0$ + 2$e$  & $k_{13}$ = $\exp$(-18.018 
		493 342 73 & \\
		& + 2.360 852 208 681 ln $T_e$&\\
		& - 2.827 443 061 704 $\times$ 10$^{-1}$ (ln $T_e$)$^2$&\\
		& + 1.623 316 639 567 $\times$ 10$^{-2}$ (ln $T_e$)$^3$&\\
		& - 3.365 012 031 362 999 $\times$ 10$^{-2}$ (ln $T_e$)$^4$&\\
		& + 1.178 329 782 711  $\times$ 10$^{-2}$ (ln $T_e$)$^5$&\\
		& - 1.656 194 699 504  $\times$ 10$^{-3}$ (ln $T_e$)$^6$&\\
		& + 1.068 275 202 678  $\times$ 10$^{-4}$ (ln $T_e$)$^7$&\\
		& - 2.631 285 809 207  $\times$ 10$^{-6}$ (ln $T_e$)$^8$& \\
		15) H$^-$ + H$^+$ $\rightarrow$ 2H$^0$ + $\gamma$ & $k_{15} = 4\times 
		10^{-8} \left({T_K}/{300}\right)^{-0.5}$ &\\
		\hline
	\end{tabular}
\end{table*}

\end{appendix}

\clearpage

\bibliographystyle{utphys}
\bibliography{mirror-ref}

\end{document}